\documentclass[12pt]{article}
\usepackage{amsfonts}
\usepackage{amssymb}
\usepackage{amsbsy}
\usepackage{mathrsfs}
\usepackage{amsmath}
\usepackage{amsthm}
\usepackage{graphicx}    
\usepackage{rotating}    
\usepackage{epsfig}
\usepackage{color}       

\input epsf.sty

\newlength{\TZ}
\newcommand{\BEQ}{\begin{equation}}     
\newcommand{\BEA}{\begin{eqnarray}}
\newcommand{\BD}{\begin{displaymath}}
\newcommand{\EEQ}{\end{equation}}       
\newcommand{\EEA}{\end{eqnarray}}
\newcommand{\ED}{\end{displaymath}}
\newcommand{\D}{{\rm d}}                
\newcommand{\II}{{\rm i}}               
\newcommand{\demi}{\frac{1}{2}}         
\newcommand{\wit}[1]{\widetilde{#1}}    
\newcommand{\wht}[1]{\widehat{#1}}      

\renewcommand{\vec}[1]{\boldsymbol{#1}} 

\newcommand{\appsektion}[1]{\setcounter{equation}{0}\setcounter{subsection}{0}
\section*{Appendix. #1}
\renewcommand{\theequation}{A.\arabic{equation}}
              \renewcommand{\thesection}{A} }

\newcommand{\slin}{{\mathfrak{sl}}}

\newcommand{\R}{\mathbb{R}}

\newcommand{\Z}{\mathbb{Z}}

\catcode`\@=11
\def\numberbysection{\@addtoreset{equation}{section}
        \def\theequation{\thesection.\arabic{equation}}}
\numberbysection

\begin{document}

\begin{titlepage}

\vskip 1.5 cm
\begin{center}
{\LARGE \bf Meta-Schr\"odinger invariance}
\end{center}

\vskip 2.0 cm
\centerline{{\bf Stoimen Stoimenov}$^a$ and {\bf Malte Henkel}$^{b,c}$}
\vskip 0.5 cm
\centerline{$^a$ Institute of Nuclear Research and Nuclear Energy, Bulgarian Academy of Sciences,}
\centerline{72 Tsarigradsko chaussee, Blvd., BG -- 1784 Sofia, Bulgaria}
\vspace{0.5cm}
\centerline{$^b$Laboratoire de Physique et Chimie Th\'eoriques (CNRS UMR 7019),}
\centerline{Universit\'e de Lorraine Nancy, B.P. 70239, F -- 54506 Vand{\oe}uvre l\`es Nancy Cedex, France}
\vspace{0.5cm}
\centerline{$^c$Centro de F\'{i}sica Te\'{o}rica e Computacional, Universidade de Lisboa,}
\centerline{Campo Grande, P--1749-016 Lisboa, Portugal}

\begin{abstract}
The Meta-Schr\"odinger algebra arises as the dynamical symmetry in transport processes which are ballistic in a chosen `parallel' direction
and diffusive in all other `transverse' directions. The time-space transformations of this Lie algebra and its infinite-dimensional extension, the
meta-Schr\"odinger-Virasoro algebra, are constructed. We also find the representations suitable for non-stationary systems by proposing a generalised
form of the generator of time-translations. Co-variant two-point functions of quasi-primary scaling operators are derived for both the
stationary and the non-stationary cases.
\end{abstract}

\end{titlepage}

\setcounter{footnote}{0}

\section{Introduction}

Dynamical symmetries often provide useful insight into strongly interacting many-body systems.
For equilibrium critical phenomena, (ortho-)conformal invariance has led to profound insights,
especially for classical systems in two spatial dimensions or for one-dimensional quantum systems, where the dynamical
symmetry algebra is isomorphic to the direct sum $\mathfrak{vir}\oplus\mathfrak{vir}$ of two Virasoro algebras
\cite{Belavin84,Francesco97,Henkel99,Rychkov17}.\footnote{In $d>2$ dimensions, the conformal Lie algebra $\mathfrak{conf}(d)$ is finite-dimensional.
In terms of the simple complex Lie algebras $B_n$, $D_n$ of Cartan's classification, one has $\mathfrak{conf}\bigl(2n\bigr)\cong D_{n+1}$ and
$\mathfrak{conf}\bigl(2n+1\bigr)\cong B_{n+1}$, with the integer $n\geq 1$.}
Here, we are interested in dynamical symmetries in non-equilibrium situations, for a review, see \cite{Bernard16}.
The first paradigmatic example for these is the free diffusion equation, where $D$ is the diffusion constant
\BEQ \label{1.1}
\mathscr{S}\Phi(t,\vec{r})=\bigl(\partial_t-D \vec{\nabla}_{\vec{r}}\cdot\vec{\nabla}_{\vec{r}} \bigr)\Phi(t,\vec{r})=0 \;\; , \;\;
\wht{\mathscr{S}}\,\wht{\Phi}(t,\vec{q}) = \bigl( \partial_t + D \vec{q}^2 \bigr)\wht{\Phi}(t,\vec{q}) = 0
\EEQ
in $1+d$ time-space dimensions and both in direct space as well as in Fourier space. The maximal dynamical symmetry Lie algebra of (\ref{1.1}) is known
since a long time \cite{Jacobi1842,Lie1881} and is usually called the {\em Schr\"odinger algebra} \cite{Niederer72,Jackiw72,Hagen72}
$\mathfrak{sch}(d)\cong \mathfrak{sl}(2,\mathbb{R})\ltimes\bigl( \mathfrak{so}(d)\ltimes\mathfrak{hei}(d)\bigr)$, written here as semi-direct
sums of special linear, rotation and Heisenberg Lie algebras. There is an infinite-dimensional extension \cite{Henkel94}, usually called
the {\em Schr\"odinger-Virasoro Lie algebra} \cite{Unterberger06,Unterberger12}.
Physically, besides the obvious examples of simple diffusion/Schr\"odinger equations \cite{Toppan15},
Schr\"odinger-invariance arises as a dynamical symmetry in non-equilibrium
dynamics such that the so-called {\em dynamical exponent} $z=2$,
which occurs for example in the ageing phenomena\footnote{Then the dynamical exponent $z$ describes the long-time behaviour of the
single relevant physical length scale (size of ordered clusters) $\ell(t)\sim t^{1/z}$ \cite{Bray94}.} of spin systems quenched to below
their critical temperature $T_c>0$ from a totally disordered initial state \cite{Bray94,Henkel10,Henkel17c}.

A second example arises if one considers the dynamics of spin systems with a directional bias. The most simple
example of this is the equation of ballistic transport, in $1+(d-1)$ spatial dimensions (again, both in direct and Fourier space)
\BEQ \label{1.2}
\mathscr{B}\Phi(t,r_{\|},\vec{r}_{\perp})=\bigl( \partial_t - S_{\|}\partial_{\|} \bigr)\Phi(t,r_{\|},\vec{r}_{\perp}) = 0 \;\; , \;\;
\wht{\mathscr{B}}\,\wht{\Phi}(t,q_{\|},\vec{q}_{\perp})=\bigl( \partial_t - \II S_{\|}q_{\|}  \bigr)\wht{\Phi}(t,q_{\|},\vec{q}_{\perp}) = 0
\EEQ
Its dynamical symmetry is the so-called {\em meta-conformal Lie algebra} $\mathfrak{meta}(1,d-1)$ \cite{Henkel10,Henkel19a}.
In $d=1$ and $d=2$ space dimensions, respectively, the Lie algebra is infinite-dimensional (and then denoted $\mathfrak{metav}(1,d-1)$), where
$\mathfrak{metav}(1,1)\cong \mathfrak{vir}\oplus\mathfrak{vir}$ and\footnote{Throughout, we consider representations of
$\mathfrak{vir}$ with vanishing central charge.\\ The Virasoro algebra with vanishing central charge
goes back at least to Cartan \cite{Cartan1909}. An often-used modern name is {\it `loop algebra'}.
Witt studied later (1935) a form of the loop algebra over fields of characteristic $p>0$, with generators $\ell_n$, $-1\leq n\leq p-2$.}
$\mathfrak{metav}(1,2)\cong \mathfrak{vir}\oplus\mathfrak{vir}\oplus\mathfrak{vir}$.
On the other hand, for $d>3$, the meta-conformal algebra is finite-dimensional. In terms of the simple complex Lie algebras
$A_1, B_n, D_n$ from Cartan's classification, e.g. \cite{Francesco97},
one expects $\mathfrak{meta}(1,2n)\cong A_1\oplus D_{n+1}$ and $\mathfrak{meta}(1,2n+1)\cong A_1\oplus B_{n+1}$ with $n=1,2,\ldots$ \cite{Henkel19a}.
Using the standard representations in terms of infinitesimal time-space transformations and integrating them, one obtains the associated
group transformations, as listed in table~\ref{tab1}. In particular, it can be seen that in two time-space dimensions, the Lie algebras of ortho-conformal
and meta-conformal transformations are isomorphic, although the physically relevant representations are different (only the ortho-conformal
time-space transformations are angle-preserving), see \cite{Henkel19a} for details on the physical choice of the abstract coordinates for
ortho- and meta-conformal transformations. Meta-conformal invariance, which has a dynamical exponent $z=1$, is realised in the relaxation dynamics
of the directed Glauber-Ising chain \cite{Godreche11,Godreche15a},
quenched from an disordered initial state to temperature $T=0$ and with long-ranged initial correlations \cite{Henkel19a}. The dynamic exponent $z=1$
also  arises in closed non-equilibrium quantum systems \cite{Calabrese16,Delfino17}.

The possibility of realising meta-conformal symmetries for spatial dimensions $d\geq 2$ raises a problem.
In momentum space, an appropriate equation of motion might look as follows
\BEQ \label{1.3}
\wht{\mathscr{B}}\,\wht{\Phi}(t,q_{\|},\vec{q}_{\perp})
=\bigl( \partial_t - \II S_{\|}q_{\|} + {\rm O}\bigl(q_{\|}^2, \vec{q}_{\perp}^2\bigr) \bigr)\wht{\Phi}(t,q_{\|},\vec{q}_{\perp}) = 0
\EEQ
In the direction of the bias, terms of order ${\rm O}\big(q_{\|}^2\bigr)$ can be viewed as merely giving rise to corrections to the long-time and
large-distance behaviour. On the other hand, terms of order ${\rm O}\big(\vec{q}_{\perp}^2\bigr)$ are {\it a priori} of leading order.
Such terms will only leads to corrections to scaling if the dynamical exponent is fixed as $z=1$ from the outset also for all transverse directions.
Otherwise, one must take into account that the dynamical exponent may take different values \cite{Schmittmann95,Taeu14},
say $z_{\|}$ and $z_{\perp}$, for spatial directions parallel
and perpendicular to the direction of the bias. In the low-momenta limit, one would obtain an equation of the form
\BEQ \label{1.4}
\wht{\mathscr{S}}\,\wht{\Phi}(t,q_{\|},\vec{q}_{\perp})
=\bigl( \partial_t - \II S_{\|}q_{\|}  + S_{\perp} \vec{q}_{\perp}^2 \bigr)\wht{\Phi}(t,q_{\|},\vec{q}_{\perp}) = 0
\EEQ
with the constants $S_{\|}$ and $S_{\perp}$ and where higher-order terms in $q_{\|},\vec{q}_{\perp}$ would merely lead to corrections to scaling.
Such anisotropies are also seen in the dynamics of driven diffusive systems, see \cite{Schmittmann95} for a classic review.
Clearly, eq.~(\ref{1.4}) has spatially anisotropic dynamical exponents, viz. $z_{\|}=1$ and $z_{\perp}=2$.
In this paper, we shall find the maximal Lie algebra
of dynamical symmetries of (\ref{1.4}), adapting the methods used in \cite{Niederer72}. Since the resulting algebra will combine aspects
of meta-conformal invariance and of Schr\"odinger-invariance, we shall call it the {\em meta-Schr\"odinger algebra} $\mathfrak{metasch}(1,d-1)$.
Table~\ref{tab1} lists known examples of infinite-dimensional time-space transformations, in terms of their defining representations.
For a classification of non-relativistic Newton-Cartan time-space, see \cite{Duval09}.
The much-studied conformal Galilean algebra, e.g. \cite{Bondi62,Negro97,Henkel03a,Barnich07,Barnich13,Bagchi17},
arises for example as a symmetry of the Pais-Uhlenbeck oscillator, e.g. \cite{Gala15,Krivonos16,Fernandez20}, or else as symmetry of certain
differential equations \cite{Aizawa15,Aizawa16,Kozyrev18,Zhang10,Stoimenov15}.
In the statistical mechanics context of this work, we rather inquire about the physical nature of co-variant two-point functions
(of quasi-primary scaling operators) to be derived.
Table~\ref{tab1} shows that depending on the symmetry algebra, either correlators or else response functions are directly predicted from the co-variance of the
Lie algebra representations.

This paper is organised as follows. In section~2, we describe the construction of the new meta-Schr\"odinger algebra and its infinite-dimensional extension, the
meta-Schr\"odinger-Virasoro algebra. This is generalised in section~3 to the case without time-translation-invariance, by proposing a modification
of the standard time-translation operator such that a modified Schr\"o\-din\-ger equation maintains the dynamical symmetry.
Co-variant two-point functions of quasi-primary scaling operators are found in section~4, both for the stationary and the non-stationary case.
We conclude in section~5. An appendix discusses admissible central extensions.

A sequel paper will discuss physical realisations of meta-Schr\"odinger symmetry in the non-equilibrium dynamics of the biased spherical model.

\begin{table}
\begin{center}\begin{tabular}{|l|lll|l|} \hline
group                    & \multicolumn{3}{l|}{coordinate transformations}                                          & co-variance \\
\hline
ortho-conformal $(1+1)D$ & $z'=f(z)$      & $\bar{z}'=\bar{z}$                        &                             & correlator  \\
                         & $z'=z$         & $\bar{z}'=\bar{f}(\bar{z})$               &                             & \\ \hline
conformal galilean       & $t'=b(t)$      & \multicolumn{2}{l|}{$\vec{r}'=\left(\D b(t)/\D t\right) \vec{r}$}       & correlator \\
                         & $t'=t$         & $\vec{r}'=\vec{r}+\vec{a}(t)$             &                             & \\
                         & $t'=t$         & $\vec{r}'=\mathscr{R}(t)\vec{r}$          &                             & \\ \hline
meta-conformal $1D$      & $u=f(u)$       & $\bar{u}'=\bar{u}$                        &                             & correlator \\
                         & $u'=u$         & $\bar{u}'=\bar{f}(\bar{u})$               &                             & \\ \hline
meta-conformal $2D$      & $\tau'=\tau$   & $w'=f(w)$                                 & $\bar{w}'=\bar{w}$          & correlator \\
                         & $\tau'=\tau$   & $w'=w$                                    & $\bar{w}'=\bar{f}(\bar{w})$ & \\
                         & $\tau'=b(\tau)$ & $w'=w$                                   & $\bar{w}'=\bar{w}$          & \\ \hline
Schr\"odinger-Virasoro   & $t'=b(t)$      & \multicolumn{2}{l|}{$\vec{r}'=\left(\D b(t)/\D t\right)^{1/2} \vec{r}$} & response \\
                         & $t'=t$         & $\vec{r}'=\vec{r}+\vec{a}(t)$             &                             & \\
                         & $t'=t$         & $\vec{r}'=\mathscr{R}(t)\vec{r}$          &                             & \\ \hline
\end{tabular}\end{center}
\caption[tab1]{{\small
Examples of infinite-dimensional groups of time-space transformations, in terms of abstract coordinate transformations.
Herein, $\mathscr{R}(t)\in\mbox{\sl SO}(d)$ is a time-dependent rotation matrix and $f,\bar{f},b,\vec{a}$
are differentiable (vector) functions of their argument.
Co-variant $n$-point functions are predicted as either correlators or responses, following from
the extension of the Cartan sub-algebra \cite{Henkel14a,Henkel15,Henkel16,Henkel20}.} \label{tab1}}
\end{table}

\section{Construction of the meta-Schr\"odinger algebra}

\noindent
{\bf Definition:} {\em The {\em meta-Schr\"odinger algebra} $\mathfrak{metasch}(1,d-1)$ acts as dynamical symmetry algebra
of the following biased evolution equation
\BEQ \label{geninveq}
\mathscr{S}\Phi(t,\vec{r})=\biggl(\partial_t-S_{\|}\partial_{\|}-\sum_{i=1}^{d-1}S_{\perp_i}\partial^2_{\vec{\perp_i}}\biggr)\Phi(t,\vec{r}) = 0 ,
\EEQ
with $\partial_{\|}=\partial_{r_{\|}}$, $\partial_{\perp_i}=\partial_{\vec{r}_{\perp_i}}$ and
$S_{\|}$, $S_{\perp_i}$ are constants. We often write $\vec{r}=(r_{\|},\vec{r}_{\perp})\in\mathbb{R}\oplus\mathbb{R}^{d-1}$.}

Here, we take the directional bias in the single $\|$-direction, all $d-1$ transverse directions ($\perp$)
are un-biased. The special case $S_{\|}=0$  reduces to a simple diffusion equation in $d-1$ transverse spatial dimensions, with the maximal symmetry
$\mathfrak{sch}(d-1)$ and the special case $S_{\perp,i}=0$ ($i=1,\ldots d-1$) reduces to a ballistic transport equation in $1$ parallel direction,
but without any transport in the $d-1$ transverse directions. The dynamical symmetry is $\mathfrak{meta}(1,d-1)$.

For notational simplicity, we shall from now on concentrate on $d=2$ and shall denote $x=r_{\|}$ the coordinate in the preferred direction and $y=r_{\perp}$
the coordinate in the transverse un-biased direction. We look for the symmetry Lie algebra $\mathfrak{metasch}(1,1)$ of the equation
\BEQ \label{inveq}
\mathscr{S}\Phi(t,x,y)=\biggl(\partial_t-S_1\partial_{x}-S_2\partial^2_{y}\biggr)\Phi(t,x,y)=0.
\EEQ
Following \cite{Niederer72}, infinitesimal symmetries of (\ref{inveq}) are written in the form
\BEQ
X=-A(t,x,y)\partial_t-B(t,x,y)\partial_x-C(t,x,y)\partial_y-D(t,x,y),\label{defsym}
\EEQ
such that the functions $A,B,C,D$, which depend on $t,x,y$, are chosen in order to satisfy
\BEQ
[\mathscr{S},X]\Phi(t,x,y)=\lambda(t,x,y)\mathscr{S}\Phi(t,x,y)\label{condsatsymm}
\EEQ
for any function $\Phi(t,x,y)$. In particular, any solution $\Phi=\Phi(t,x,y)$ of eq.~(\ref{inveq})
will be mapped onto another solution of the same equation \cite{Niederer72}.
With the help of the following auxiliary identity $\partial^2_y(A\partial_i)=A_{yy}\partial_i+2A_y\partial_y\partial_i+A\partial^2_y\partial_i$,
the functions $A,B,C,D$ obey the system of equations\footnote{We use the notations $\dot{A}=\partial_tA(t,x,y)$, $A_{xy}=\partial_x\partial_yA(t,x,y)$ etc.}
\BEA
&& S_2A_{yy}+S_1A_x-\dot{A}-\lambda=0\nonumber\\
&& S_2B_{yy}+S_1B_x-\dot{B}+\lambda S_1=0\nonumber\\
&& 2S_2C_y+\lambda S_2=0\nonumber\\
&& S_2C_{yy}+S_1C_x-\dot{C}+2S_2D_y=0\label{gensystem}\\
&& S_2D_{yy}+S_1D_x-\dot{D}=0\nonumber\\
&& 2S_2A_y=0\nonumber\\
&& 2S_2B_y=0\nonumber
\EEA
Solving (\ref{gensystem}) in general is straightforward, but lengthy.
Moreover, we look for the entire symmetry algebra, that is the set of general solutions of (\ref{gensystem}) which closes into a Lie algebra.
 
As a first simplification, we note from the two last conditions that
$A=A(t,x)$ and $B=B(t,x)$. This suggests that the symmetries in the $x$- and in $y$-directions can be partially separated.
Indeed, if we fix $y=\mbox{\rm cste.}$ in (\ref{condsatsymm}), the invariant equation (\ref{inveq}) becomes
the ballistic transport equation for the spatial $x$-direction, whose known maximal symmetry is the meta-conformal algebra $\mathfrak{meta}(1,1)$. 
This is obvious from the system (\ref{gensystem}), see \cite{Henkel19a}.
On the other hand, if we fix $x=\mbox{\rm cste.}$ in (\ref{condsatsymm}), the invariant equation (\ref{inveq}) becomes
a diffusion equation whose maximal symmetry algebra is the Schr\"odinger algebra $\mathfrak{sch}(1)$. These two algebras should be obtained as
sub-algebras of the sought Lie algebra $\mathfrak{metasch}(1,1)$.
Hence a natural starting point for the general construction of $\mathfrak{metasch}(1,1)$
will be representations of the meta-conformal algebra
which leave invariant the equation
(\ref{condsatsymm}).\footnote{Alternatively, one may start from a representation of Schr\"odinger algebra which leaves invariant (\ref{inveq}).}
Recalling the algebraic structure of the meta-conformal and Schr\"odinger algebras
\BEQ\label{structure}
\mathfrak{meta}(1,1)\cong \mathfrak{sl}(2,\R)\oplus\mathfrak{sl}(2,\R), \quad
\mathfrak{sch}(1)\cong \slin(2,\R)\ltimes \mathfrak{hei}(1)
\EEQ
(it is understood that the Heisenberg algebra $\mathfrak{hei}(1)$ includes the central extension) we begin with the following

\noindent
{\bf Ansatz:} {\em A finite-dimensional representation of $\mathfrak{metasch}(1,1)$ is given
by a semi-direct sum of representations, namely of the meta-conformal algebra $\mathfrak{meta}(1,1)$, acting in the spatial direction $x$ and
the Heisenberg algebra $\mathfrak{hei}(1)$, acting in the spatial direction $y$.}\\
Clearly, for our purposes, the representations of both the meta-conformal and the Heisenberg algebras, will act on time and the two spatial coordinates.
In terms of generators this leads to
\BEA
    X_{-1} & = & -\partial_t, \quad Y_{-1}^x=-\partial_x, \quad Y^y_{-\demi}=-\partial_y, \quad M_0=-{\cal M}\nonumber\\
       X_0 & = & -t\partial_t-x\partial_x-C^{X_0}(t,x,y)\partial_y-D^{X_0}(t,x,y)-\delta\nonumber\\
       X_1 & = & -(t^2+\alpha x^2)\partial_t-(2tx+\beta x^2)\partial_x-C^{X_1}(t,x,y)\partial_y-D^{X_1}(t,x,y)-2\delta t-2\gamma x\nonumber\\
      Y_0^x & = & -\alpha x\partial_t-(t+\beta x)\partial_x-C^{Y_0}(t,x,y)\partial_y-D^{Y_0}(t,x,y)-\gamma\label{modmetasch}\\
      Y_1^x & = & -\alpha(2tx+\beta x^2)\partial_t-((t+\beta x)^2+\alpha x^2)\partial_x-C^{Y_1}(t,x,y)\partial_y-D^{Y_1}(t,x,y)\nonumber\\
            &   & -2\gamma t-2(\alpha\delta+\beta\gamma)x\nonumber\\
Y^y_{\demi} & = & -A^{Y_{\demi}}(t,x)\partial_t-B^{Y_{\demi}}(t,x)\partial_x-C^{Y_{\demi}}(t,x,y)\partial_y-D^{Y_{\demi}}(t,x,y)-{\cal M}y\nonumber
\EEA
where $\alpha,\beta$ are constants. The four unknown functions $A,B,C,D$ of each generator are obtained by using the expected sub-algebras:
\begin{itemize}
\item the meta-conformal sub-algebra $\mathfrak{meta}(1,1)=\bigl\langle X_{0,\pm 1}, Y_{0,\pm 1}^x\bigr\rangle$ with the respective
commutators (where $\alpha,\beta$ are constants)
\BEA
&& [X_n, X_m]=(n-m)X_{n+m}, \quad [X_n, Y^x_m]=(n-m)Y^x_{n+m}\nonumber\\
&& [Y^x_n, Y^x_m]=(n-m)\bigl(\alpha X_{n+m}+\beta Y^x_{n+m}\bigr)\label{metahybrid}
\EEA
The infinite-dimensional extension is $\mathfrak{metav}(1,1)=\bigl\langle X_{n}, Y_{n}^x\bigr\rangle_{n\in\mathbb{Z}}$, with
the commutators (\ref{metahybrid}) extended to $n,m\in\mathbb{Z}$  and maximal finite-dimensional sub-algebra $\mathfrak{meta}(1,1)$.
\item the Schr\"odinger sub-algebra $\mathfrak{sch}(1)=\bigl\langle  X_{0,\pm 1}, Y_{\pm\demi}^y, M_0\bigr\rangle$. For $n,m\in\mathbb{Z}$ and
$p,q\in\mathbb{Z}+\demi$, the respective non-vanishing commutators of the Schr\"odinger-Virasoro algebra $\mathfrak{sv}(1)$ read
\BEA
{}[X_n, X_m]&=&(n-m)X_{n+m}, \quad [X_n,Y^y_p]=\left(\frac{n}{2}-p\right)Y^y_{n+p}, \nonumber \\
{}[Y^y_{p}, Y^y_{q}]&=&  (p-q)M_{p+q},\quad [X_n, M_m]-m M_{n+m} \label{schhybrid}
\EEA
The maximal finite-dimensional sub-algebra of (\ref{schhybrid}) is $\mathfrak{sch}(1)$.
\end{itemize}
We begin by looking for a realisation of $\mathfrak{sl}(2,\R)=\bigl\langle X_{0,\pm 1}\bigr\rangle$
as common sub-algebra of meta-conformal and Schr\"odinger algebras.
In addition, we must check that the commutator $[Y^x_n, Y^y_p]$ closes into the algebra.
All unknown functions in the generators (\ref{modmetasch}) are found from the above commutation relations (\ref{metahybrid}, \ref{schhybrid})
and the equations (\ref{gensystem}).

Since we require time- and space-translation-invariance to hold, the meta-Schr\"odinger algebra under construction should include the
translation generators $X_{-1}=-\partial_t$, $Y_{-1}^X=-\partial_x$ and $Y_{-1/2}^y=-\partial_y$. From (\ref{modmetasch}), it follows that
the generator of dynamical scaling must be
\BEQ
X_0=-t\partial_t-x\partial_x-\frac{y}{2}\partial_y-\delta.\label{dynscaling}
\EEQ
However, several possibilities exist for the form of $X_1$. Namely, the first equation of the system (\ref{gensystem})
gives $\lambda^{X_1}=-2t+2\alpha S_1x$. Upon substitution into the second equation (\ref{gensystem}), this leads to a quadratic equation for $S_1$
\BEQ
\alpha S_1^2+\beta S_1-1=0.\label{chareq}
\EEQ
In what follows, it will be important to distinguish the cases (A) $\alpha=0$ and (B) $\alpha\ne 0$.

\subsection{Case A: {}$\alpha=0, \quad S_1=1/\beta, \quad \lambda^{X_1}=-2t$}
The representation of $\mathfrak{metasch}(1,1)$ is (where $\delta,\gamma/\beta$ act as scaling dimensions)
\BEA
X_{-1} & = & -\partial_t, \quad X_0=-t\partial_t-x\partial_x-\frac{y}{2}\partial_y-\delta\nonumber\\
X_1 & = & -t^2\partial_t-(2tx+\beta x^2)\partial_x-ty\partial_y-2t\delta-2\gamma x-\frac{\cal M}{2}y^2\nonumber\\
Y^x_{-1} & = & -\partial_x, \quad Y^x_0=-(t+\beta x)\partial_x-\gamma\nonumber\\
Y^x_1 & = & -(t+\beta x)^2\partial_x-2\gamma t-2\gamma\beta x \label{Ametasch}\\
Y^y_{-\demi} & = & -\partial_y, \quad Y^y_{\demi}=-t\partial_y-{\cal M}y, \quad M_0=-{\cal M}. \nonumber
\EEA
With $n,m\in\{\pm 1,0\}$ and $p\in\{\pm\frac{1}{2}\}$, the non-vanishing commutators are
\BEA
&& [X_n, X_m]=(n-m)X_{n+m}, \quad [X_n, Y^x_m]=(n-m)Y^x_{n+m}\nonumber\\
&& [X_n,Y^y_p]=\bigl(\frac{n}{2}-p\bigr)Y^y_{n+p}, \quad \hspace{0.35truecm}[Y^x_n, Y^x_m]=(n-m)\beta Y^x_{n+m}\label{Acomhybrid}\\
&& [Y^x_n, Y^y_p]=0, \quad \hspace{2.4truecm} [Y^y_{1/2}, Y^y_{-1/2}]= M_0.\nonumber
\EEA
In order to verify under which conditions the generators (\ref{Acomhybrid}) are indeed a dynamical
symmetry of the equation (\ref{inveq}), we consider the commutators (with $n=\pm 1,0$ and $p=\pm \frac{1}{2}$)
\BEA
&& [\mathscr{S},X_{-1}]=[\mathscr{S},Y^x_{n}]=[\mathscr{S},Y^y_{p}]=[\mathscr{S},M]=0\nonumber\\
&& [\mathscr{S},X_0]=-\mathscr{S}, \quad  \hspace{0.7truecm}
[\mathscr{S},X_1]=-2t\mathscr{S} +\bigl(2{\cal M}S_2 -1\bigr)y\partial_y -2\bigl(\delta-\frac{\gamma}{\beta}-\frac{1}{4}\bigr)\label{Asymmetryhybrid}
\EEA
which shows that a dynamical symmetry holds true if we set $S_2=\frac{1}{2{\cal M}}$ and $\frac{\gamma}{\beta}=\bigl(\delta-1/4\bigr)$.
Hence the Schr\"odinger operator is $\mathscr{S}=\partial_t - \frac{1}{\beta}\partial_x - \frac{1}{2{\cal M}}\partial_y^2$ and depends on the
non-universal parameters $\beta$ and $\cal M$. On the other hand, the universal constants $\delta$ and $\frac{\gamma}{\beta}$ describe the
transformation of scaling operators under the representation (\ref{Ametasch}) of $\mathfrak{metasch}(1,1)$.
The generalisation of the representation (\ref{Ametasch}) of the finite-dimensional meta-Schr\"odinger algebra
to the infinite-dimensional {\em meta-Schr\"odinger-Virasoro algebra} $\mathfrak{msv}(1,1)$ reads, with  $n\in\mathbb{Z}$ and $p\in\Z+\demi$
\BEA
X_n & = & -t^{n+1}\partial_t-\frac{1}{\beta}\left((t+\beta x)^{n+1}-t^{n+1}\right)\partial_x\nonumber\\
    &   & -(n+1)\left(\demi t^ny\partial_y+\frac{\gamma}{\beta}\bigl((t+\beta x)^n-t^n\bigr)+t^n\delta+\frac{n}{4}{\cal M}t^{n-1}y^2\right)\nonumber\\
Y^x_n & = & -(t+\beta x)^{n+1}\partial_x-(n+1)\gamma (t+\beta x)^n\nonumber\\
Y^y_p & = & -t^{p+\demi}\partial_y-\bigl(p+\demi\bigr){\cal M}t^{p-\demi}y, \nonumber \\
M_n &=& -{\cal M}t^n.\label{Ainfinitty}
\EEA
and whose non-vanishing commutators are given by the extension of (\ref{Acomhybrid}). In view of the explicit form (\ref{Ainfinitty}), it is easily seen that
the natural variables are not $t,x,y$ but rather $\tau=t,v=t+\beta x,y$. In these variables, the meta-Schr\"odinger-Virasoro algebra is spanned by the generators
\BEA
A_n & := & X_n -\frac{1}{\beta}Y_n^x
      = -{\tau}^{n+1}\partial_{\tau}-(n+1)\left(\demi \tau^n y\partial_y+\bigl(\delta-\frac{\gamma}{\beta}\bigr)\tau^n+\frac{n}{4}{\cal M}\tau^{n-1}y^2\right)\nonumber\\
Y^x_n & = & -\beta v^{n+1}\partial_v-(n+1)\gamma v^n,\nonumber \\
Y^y_p &=& -\tau^{p+\demi}\partial_y-(p+\demi){\cal M}\tau^{p-\demi}y, \nonumber \\
M_n &=&-{\cal M}\tau^n,\label{2Ainfinitty}
\EEA
whose non-vanishing commutators ($n,m\in\mathbb{Z}$ and $p,q\in\mathbb{Z}+\frac{1}{2}$)
\BEA
&& [A_n, A_m]=(n-m)A_{n+m}, \quad\hspace{0.4cm} [A_n,Y^y_p]=\bigl(\frac{n}{2}-p\bigr)Y^y_{n+p},\nonumber \\
&& [Y^x_n, Y^x_m]=(n-m)\beta Y^x_{n+m}\quad \hspace{0.5truecm} [Y^y_{p}, Y^y_{q}]= (p-q)M_{p+q} \label{Acomhybrid2}
\EEA
make evident the Lie algebra structure
\BEQ
\mathfrak{msv}(1,1)= \mathfrak{vir}_v\oplus\bigl(\mathfrak{vir}_{\tau}\ltimes\mathfrak{gal}_{\tau y}(1)\bigr)=
                            \mathfrak{vir}_v\oplus\mathfrak{sv}_{\tau y}(1).\label{Astructure}
\EEQ
Herein, $\mathfrak{gal}(1)=\bigl\langle Y_p^{y}, M_n\bigr\rangle_{p\in\mathbb{Z}+\frac{1}{2},n\in\mathbb{Z}}$ is the
infinite-dimensional extension of the Heisenberg algebra $\mathfrak{hei}(1)$,
$\mathfrak{vir}$ is the Virasoro algebra, $\mathfrak{sv}(1)$ is the Schr\"odinger-Virasoro algebra,
and the subscript indices label the explicit variables in (\ref{Ainfinitty}).
In the variables $(\tau,v,y)=(t,t+\beta x,y)$, the Schr\"odinger operator (\ref{inveq}) simplifies into
\BEQ\label{Ainveq}
\mathscr{S} = \partial_t -\frac{1}{\beta}\partial_x - \frac{1}{2{\cal M}}\partial_y^2 = \partial_{\tau}-\frac{1}{2{\cal M}}\partial_y^2
\EEQ
Hence the dynamical symmetry of the equation (\ref{inveq}) is the algebra $\mathfrak{vir}_v\oplus\mathfrak{sch}_{ty}(1)$
(which does not include the full Schr\"odinger-Virasoro algebra). This is natural because the hidden variable $v=t+\beta x$ does not occur in eq.~(\ref{Ainveq}).

We point out that in the representation (\ref{2Ainfinitty}), in the reduced generator $A_n=X_n -\frac{1}{\beta}Y_n^x$ expressed in terms of light-cone coordinates
$\tau,v$, the effective scaling dimension now reads $\delta-\frac{\gamma}{\beta}$ instead of $\delta$. We shall see that this feature arises systematically.

\subsection{Case B:  $\alpha\ne 0, \quad \lambda^{X_1}=-2t+2\alpha S_1x$ }
We  set $c:=-\alpha S_1$. From (\ref{chareq}), we then have $\alpha=c(c-\beta)\ne 0$.
The representation is
\BEA
X_{-1} & = & -\partial_t, \quad X_0=-t\partial_t-x\partial_x-\frac{y}{2}\partial_y-\delta\nonumber\\
X_1 & = & -(t^2+\alpha x^2)\partial_t-(2tx+\beta x^2)\partial_x-(t+cx)y\partial_y-2t\delta-2\gamma x-\frac{\cal M}{2}y^2\nonumber\\
Y^x_{-1} & = & -\partial_x, \quad Y^x_0=-\alpha x\partial_t-(t+\beta x)\partial_x-\frac{c}{2}y\partial_y-\gamma\nonumber\\
Y^x_1 & = & -\alpha(2tx+\beta x^2)\partial_t-(t^2+2\beta tx+(\alpha+\beta^2)x^2)\partial_x-(ct+(\alpha+\beta c)x)y\partial_y\nonumber\\
      &   & -2\gamma t-2(\alpha\delta+\beta \gamma)x-\frac{c{\cal M}}{2}y^2\nonumber\\
Y^y_{\demi} & = & -\partial_y, \quad Y^y_{\demi}=-(t+cx)\partial_y-{\cal M}y, \quad M_0=-{\cal M}.\label{Bmetasch}
\EEA
The generators (\ref{Bmetasch}) satisfy the following non-vanishing commutation relations, with $n,m=\pm 1,0$ and $p=\pm \demi $
\BEA
&& [X_n, X_m]=(n-m)X_{n+m}, \quad [X_n, Y^x_m]=(n-m)Y^x_{n+m}\nonumber\\
&& [X_n,Y^y_p]=\bigl(\frac{n}{2}-p\bigr)Y^y_{n+p}, \quad \hspace{0.3truecm}[Y^x_n, Y^x_m]=(n-m)\bigl(\alpha X_{n+m}+\beta Y^x_{n+m}\bigr)\nonumber\\
&& [Y^x_n, Y^y_p]=c\bigl(\frac{n}{2}-p\bigr)Y^y_{n+p}, \quad \hspace{0.1truecm}[Y^y_{1/2}, Y^y_{-1/2}]= M_0.\label{Bcomhybrid}
\EEA
Next, from the above we have $S_1=-\frac{c}{\alpha}=-\frac{1}{c-\beta}$. In order to check the dynamical invariance of eq.~(\ref{inveq}) under the
representation (\ref{Bmetasch}), we consider the commutators
\BEA
&& [\mathscr{S},X_{-1}]=[\mathscr{S},Y^x_{-1}]=[\mathscr{S},Y^y_{-\demi}]=[\mathscr{S},Y^y_{\demi}]=[\mathscr{S},M_0]=0\nonumber\\
&& [\mathscr{S},X_0]=-\mathscr{S}, \quad [\mathscr{S},Y^x_0]=-c\mathscr{S} \label{symmetry}\\
&& [\mathscr{S},Y^x_1]=-2c(t+cx)\mathscr{S}+c\bigl(2{\cal M}S_2 - \frac{2c-\beta}{c-\beta}\bigr)y\partial_y
   +\frac{2c}{\beta-c}\bigl(\gamma-(\beta-c)\delta-\frac{2c-\beta}{4}\bigr),\nonumber\\
&& [\mathscr{S},X_1]=-2(t+cx)\mathscr{S}+\bigl(2{\cal M}S_2 - \frac{2c-\beta}{c-\beta}\bigr)y\partial_y
   +\frac{2}{\beta-c}\bigl(\gamma-(\beta-c)\delta-\frac{2c-\beta}{4}\bigr),\nonumber
\EEA
from which we read off the conditions (distinct from those for the case $\alpha=0$)
\BEQ \label{symm-cond}
S_2=\frac{1}{2{\cal M}}\frac{2c-\beta}{c-\beta} \quad \mbox{\rm\small and} \quad \gamma= \frac{2c-\beta}{4}+(\beta-c)\delta
\EEQ
for the dynamical symmetry.  The Schr\"odinger operator has the final form
$\mathscr{S}=\partial_t +\frac{1}{c-\beta}\partial_x -\frac{1}{2{\cal M}}\frac{2c-\beta}{c-\beta}\partial_y^2$. \\

In order to obtain infinite-dimensional extensions of the representation (\ref{Bmetasch}) of the meta-Schr\"odinger algebra, we note
that the infinite-dimensional extension of the Heisenberg algebra can be done in terms of the variable $\rho=t+cx$
and reads
\BEQ
Y_p^y=-\rho^{p+1/2}\partial_y-(p+1/2){\cal M}\rho^{p-1/2}y \;\; ~,~ \;\;  M_n=-{\cal M}\rho^n\label{Bgalileiinf}
\EEQ
such that for $p,q\in\Z+\demi$ we have the (only non-vanishing) commutator
\BEQ
[Y_p^y,Y_q^y]=(p-q)M_{p+q}.\label{comGal}
\EEQ
Next, following \cite{Henkel19a}, we look for a new family of generators, first for $n=0,\pm 1$, in the form
\BEQ
{\cal Y}_n=\mathscr{N}\,\big(a X_n+Y_n^x\bigr)\label{calydef}
\EEQ
where the normalisation $\mathscr{N}$ will be fixed shortly. The new generators satisfy
\BEQ
[{\cal Y}_n,{\cal Y}_m]=(n-m)(2a+\beta)\mathscr{N}{\cal Y}_{n+m}\label{comcaly}
\EEQ
provided that $a$ satisfies $a^2+\beta a-c(c-\beta)=0$.
The two solutions $a_{1,2}$ of this quadratic equation, namely $a_1=-c$ and $a_2=c-\beta$,
produce two distinct forms, denoted ${\cal Y}^{(1,2)}$, of the generator. We then obtain
\BEA
     {\cal Y}^{(1)}_n & = & \mathscr{N}^{(1)}\bigl(-c X_n+Y_n^x\bigr), \quad \hspace{0.8truecm}
     [{\cal Y}_n^{(1)},{\cal Y}_m^{(1)}]=(n-m)(\beta-2c)\mathscr{N}^{(1)}{\cal Y}_{n+m}^{(1)}\nonumber\\
     {\cal Y}^{(2)}_n & = & \mathscr{N}^{(2)}\bigl((c-\beta)X_n+Y_n^x\bigr), \quad
     [{\cal Y}_n^{(2)},{\cal Y}_m^{(2)}]=(n-m)(2c-\beta)\mathscr{N}^{(2)}{\cal Y}_{n+m}^{(2)}\label{calyresults}
\EEA
which both satisfy the commutator (\ref{comcaly}). We now fix the normalisations from the requirements
$(\beta-2c)\mathscr{N}^{(1)}=(2c-\beta)\mathscr{N}^{(2)}\stackrel{!}{=}1$.

Analogously, we construct new generators
\BEQ
{\cal A}_n=X_n+b{\cal Y}_n\label{calAdef}
\EEQ
and require that they should obey the commutator
\BEQ
[{\cal A}_n, {\cal A}_m]=(n-m){\cal A}_{n+m},\label{comcalA}
\EEQ
In the case where $a=a_1=-c$, we have ${\cal A}^{(1)}=X_n+b{\cal Y}_n^{(1)}$ where $b$ must satisfy
\BEQ
\bigl(b\mathscr{N}^{(1)}\bigr)^2(\beta-2c)+b\mathscr{N}^{(1)} =0 \quad \Longrightarrow \quad
\left\{ \begin{array}{l} b_0=0 \\  b_1=-1 \end{array}\right. \label{detAn1}
\EEQ
In the case where $a=a_2=c-\beta$, we have ${\cal A}^{(2)}=X_n+b{\cal Y}_n^{(2)}$ where $b$ must satisfy
\BEQ
\bigl(b\mathscr{N}^{(2)}\bigr)^2(\beta-2c)-b\mathscr{N}^{(2)} =0  \quad \Longrightarrow \quad
\left\{ \begin{array}{l} b_0=0 \\  b_1=-1 \end{array}\right. \label{detAn2}
\EEQ
Hence three distinct forms of the ${\cal A}_n$ are possible (one for $b=0$ and two for $b=-1$), namely
\BEA
{\cal A}^{(0)}_n&=&X_n, \nonumber \\
{\cal A}^{(1)}_n &=&X_n-{\cal Y}_n^{(1)} = \frac{(c-\beta)X_n + Y_{n}^x}{2c-\beta} = {\cal Y}_n^{(2)} \label{calAresults} \\
{\cal A}^{(2)}_n &=&X_n-{\cal Y}_n^{(2)} = -\frac{-cX_n + Y_{n}^x}{2c-\beta} \hspace{0.4cm}= {\cal Y}_n^{(1)} \nonumber
\EEA

In addition: if $b=0$ we find $[{\cal A}_n,{\cal Y}_m]=(n-m){\cal Y}_{n+m}$ but if $b=-1$, we find $[{\cal A}_n,{\cal Y}^{(1,2)}_m]=0$.

The construction presented here is valid if $\beta\ne 2c$ and
$\beta\ne c$.\footnote{For the limit case $\beta=2c$, a Lie algebra contraction should lead to representations
related to the conformal Galilean algebra.} Then the maximal finite-dimensional sub-algebra
(\ref{Bmetasch}) is the dynamical symmetry of the equation (\ref{inveq}).

We now write down the generators explicitly for the admissible values of $b$.

\subsubsection{$b=-1$}
Because of (\ref{calAresults}), we have ${\cal A}_n^{(1)}={\cal Y}_n^{(2)}$, and ${\cal A}_n^{(2)}={\cal Y}_n^{(1)}$. Hence, these possibilities are
not independent. Let ${\cal A}^{B}_n={\cal A}^{(1)}_n$ and ${\cal Y}^{B}_n={\cal Y}^{(1)}_n$.
The explicit expressions for $X_n, Y_n^x$ in (\ref{Bmetasch}) give
\BEA
{\cal Y}^B_{-1} & = & -\frac{1}{2c-\beta}\bigl( c\partial_t-\partial_x\bigr)\nonumber\\
   {\cal Y}^B_0 & = & -\frac{1}{2c-\beta}\bigl((t+(\beta-c)x)(c\partial_t-\partial_x)+c\delta-\gamma\bigr) \label{finitcaly}\\
   {\cal Y}^B_1 & = & -\frac{1}{2c-\beta}\bigl((t+(\beta-c)x)^2(c\partial_t-\partial_x)+2(c\delta-\gamma)(t+(\beta-c)x)\bigr) \nonumber
\EEA
This result is readily extended to its infinite-dimensional form ($n\in\mathbb{Z}$)
\BEQ \label{Bcalyafinal}
{\cal Y}^B_n=-\frac{1}{2c-\beta}\Bigl(\bigl(t+(\beta-c)x\bigr)^{n+1}(c\partial_t-\partial_x)+(n+1)(c\delta-\gamma)\bigl(t+(\beta-c)x\bigr)^n\Bigr)
\EEQ
Next, for the generators $A^B_n$ we have
\BEA
{\cal A}^B_{-1} & = & -\frac{1}{2c-\beta}\Bigl((c-\beta)\partial_t+\partial_x\Bigr) \nonumber\\
{\cal A}^B_0    & = & -\frac{1}{2c-\beta}\Bigl((t+cx)\bigl((c-\beta)\partial_t+\partial_x\bigr)+(c-\beta)\delta+\gamma\Bigr)-\frac{y}{2}\partial_y \label{finitA}\\
{\cal A}^B_1    & = & -\frac{1}{2c-\beta}\Bigl((t+cx)^2\bigl((c-\beta)\partial_t+\partial_x)+2((c-\beta)\delta+\gamma\bigr)(t+cx)\Bigr)
                      -(t+cx)y\partial_y-\frac{{\cal M}}{2}y^2\nonumber
\EEA
with the infinite-dimensional extension
\BEA
{\cal A}^B_n & = & -\frac{1}{2c-\beta}\Bigl((t+cx)^{n+1}((c-\beta)\partial_t+\partial_x)+(n+1)((c-\beta)\delta+\gamma)(t+cx)^n\Bigr)\nonumber\\
             &   & -\frac{(n+1)}{2}(t+cx)^ny\partial_y-\frac{n(n+1)}{4}(t+cx)^{n-1}{\cal M}y^2\label{calBafinal}
\EEA
Finally, working with new light-cone-like variables
\BEQ \label{cone-lumiere}
\sigma=t+(\beta-c)x \;\; , \;\; \rho=t+cx
\EEQ
the generators  ${\cal Y}^B_n$ (\ref{Bcalyafinal}) and ${\cal A}^B_n$ (\ref{calBafinal}) become
\begin{subequations} \label{calAY-B}
\begin{align}
{\cal Y}^{B}_n & = -\sigma^{n+1}\partial_{\sigma}-(n+1)\frac{c\delta-\gamma}{2c-\beta}\sigma^n \label{calYsigma}\\
{\cal A}^{B}_n & = -\rho^{n+1}\partial_{\rho}-(n+1)\left(\frac{(c-\beta)\delta+\gamma}{2c-\beta}+\frac{y}{2}\partial_y\right)\rho^n
                     -\frac{n(n+1)}{4}{\cal M}y^2\rho^{n-1}\label{calBrho}
\end{align}
\end{subequations}
and the list of generators is completed by (\ref{Bgalileiinf}).
The non-vanishing commutators are, for $n,m\in\mathbb{Z}$ and $p,q\in\mathbb{Z}+\demi$
\BEA
&& [{\cal A}^B_n, {\cal A}^B_m]=(n-m){\cal A}^B_{n+m}, \quad [{\cal Y}^B_n, {\cal Y}^B_m]=(n-m){\cal Y}^B_{n+m}\nonumber\\
&& [{\cal A}^B_n,Y^y_p]=\left(\frac{n}{2}-p\right)Y^y_{n+p},\quad~  [Y^y_p, Y^y_q]=(p-q)M_{p+q},  \quad [{\cal A}^B_n,M_m]=-m M_{n+m}~~
\label{msv-B}
\EEA
which reproduces the meta-Schr\"odinger-Virasoro Lie algebra (\ref{Astructure})
$\mathfrak{msv}(1,1)\cong \mathfrak{vir}_{\sigma}\oplus\mathfrak{sv}(1)_{\rho y}$.
Non-vanishing central extensions (of course of the familiar Virasoro form) only exist for the commutators
$[{\cal A}^B_n, {\cal A}^B_m]$ and $[{\cal Y}^B_n, {\cal Y}^B_m]$, as shown in the appendix.

\subsubsection{$b=0$}

Fixing the generators ${\cal A}_n^{(0)}=X_n$, there are two possible realisations of ${\cal Y}^{(1,2)}_n$,
see eqs.~(\ref{calyresults},\ref{calAresults}). Consider first the representation spanned by $\{X_n,{\cal Y}^{(1)}_n\}$, where
${\cal Y}^{(1)}_n ={\cal Y}^B_n$ is given by (\ref{calYsigma}). Again, use the variables $\sigma,\rho$ given in (\ref{cone-lumiere}). For example,
\BEQ
X_1=-\rho^2\partial_{\rho}-\sigma^2\partial_{\sigma}-\rho y\partial_y
    -{2}\left(\frac{(c-\beta)\delta+\gamma}{2c-\beta}\rho+\frac{c\delta-\gamma}{2c-\beta}\sigma\right)-\frac{{\cal M}}{2}y^2.\label{X1general}
\EEQ
Taking into account the explicit form of ${\cal Y}^{B}_n$ (\ref{calYsigma}), it is useful to decompose $X_1 = {\cal Y}^B_1 + {\cal X}_1$.
In general, for $n\in\mathbb{Z}$ we have similarly $X_n = {\cal Y}^B_n + {\cal X}_n$, where
\BEQ
{\cal X}_n  =  X_n - {\cal Y}^B_n =
                 -\rho^{n+1}\partial_{\rho}-\frac{n+1}{2}\rho^n y\partial_y
                 -{(n+1)}\frac{(c-\beta) \delta+\gamma}{2c-\beta}\rho^n-\frac{n(n+1)}{4}{\cal M}\rho^{n-1}y^2~~ \label{calXn}
\EEQ
and from (\ref{calBrho}), we read off ${\cal X}_n={\cal A}^B_n$. Therefore, the representations  $\{X_n,{\cal Y}^{(1)}_n \}$ and
$\{{\cal A}^B_n, {\cal Y}^B_n\}$ are isomorphic, up to a change of basis. Explicitly
\BEA
&& {\cal X}_n = - \rho^{n+1}\partial_{\rho} - \frac{n+1}{2}\rho^n y\partial_y - (n+1)(\Delta-\Gamma)\rho^n - \frac{n(n+1)}{4}{\cal M} \rho^{n-1} y^2 \nonumber \\
&& {\cal Y}^B_n = -\sigma^{n+1}\partial_{\sigma} - (n+1) \Gamma \sigma^n \label{gen-casB}\\
&& Y_p^y = - \rho^{p+1/2}\partial_y - \bigl(p+\frac{1}{2}\bigr) {\cal M} \rho^{p-1/2} y \nonumber \\
&& M_n = - {\cal M}\rho^n \nonumber
\EEA
in terms of the non-orthogonal light-cone coordinates (\ref{cone-lumiere}), with the abbreviations
\BEQ \label{defGammaDelta}
\Gamma := \frac{c\delta - \gamma}{2c-\beta} \;\; , \;\; \Delta - \Gamma := \frac{(c-\beta)\delta +\gamma}{2c-\beta} ~~\Longrightarrow \Delta = \delta
\EEQ
These are motivated by the representation (\ref{2Ainfinitty}), which shows that when working in light-cone coordinates, the effective scaling dimension $\delta$
in the reduced generator $A_n$ is changed into $\delta-\gamma$ (if one uses the scaling $\beta=1$). We apply the same interpretation here to the reduced
generator ${\cal X}_n$.
The non-vanishing commutators are, written down here using ${\cal A}^{(0)}_n = X_n$
\BEA
&& [{\cal A}^{(0)}_n, {\cal A}^{(0)}_m]=(n-m){\cal A}^{(0)}_{n+m},
   \quad [{\cal A}^{(0)}_n, {\cal Y}^{B}_m]=(n-m){\cal Y}^B_{n+m}, \quad [{\cal Y}^B_n, {\cal Y}^B_m]=(n-m){\cal Y}^B_{n+m}\nonumber\\
&& [{\cal A}^{(0)}_n,Y^y_p]=\left(\frac{n}{2}-p\right)Y^y_{n+p},\quad~~  [Y^y_p, Y^y_q]=(p-q)M_{p+q},  \quad~~~ [{\cal A}^{(0)}_n,M_m]=-m M_{n+m}~~~~~~
\EEA
while if we had used ${\cal X}_n$ instead of ${\cal A}^{(0)}_n$, the structure would be reduce to (\ref{msv-B}).
An analogous consideration with the generators ${\cal Y}^{(2)}_n$ leads to the same conclusion. Therefore,
all possible infinite-dimensional realisations of the algebra in the case {\bf B} are identical or equivalent to the generators (\ref{calAY-B}).

The results of this section are summarised as follows.

\noindent
{\bf Proposition:} {\em The meta-Schr\"odinger-Virasoro algebra
$\mathfrak{msv}(1,1)=\left\langle {\cal A}_n^B,{\cal Y}^B_n,Y_p^y,M_n\right\rangle_{n\in\mathbb{Z},p\in\mathbb{Z}+\frac{1}{2}}$
is a semi-direct sum of the $1D$ meta-conformal algebra and the infinite-dimensional extension $\mathfrak{gal}(1)$
of the Heisenberg algebra\footnote{The indices refer to the coordinates used in (\ref{calAY-B},\ref{gen-casB}).}
\BEQ \label{Bstructure}
\mathfrak{msv}(1,1)= \bigl(\mathfrak{vir}_{\sigma}\oplus\mathfrak{vir}_{\rho}\bigr)\ltimes\mathfrak{gal}_{\rho y}(1)
                   =       \mathfrak{vir}_{\sigma}\oplus\mathfrak{sv}_{\rho y}(1),
\EEQ
with commutators given by (\ref{msv-B}).
A sub-algebra of $\mathfrak{msv}(1,1)$ is the Schr\"odinger-Virasoro algebra
$\mathfrak{sv}(1)=\left\langle {\cal A}_n^B,Y_p^y,M_n\right\rangle_{n\in\mathbb{Z},p\in\mathbb{Z}+\frac{1}{2}}$.
Central extensions only arise in the sub-algebras $\mathfrak{vir}$ and $\mathfrak{sv}(1)$ (they are not spelled out in the above commutators).

However, only the maximal sub-algebra $\mathfrak{vir}\oplus\mathfrak{sch}(1)$ acts as a dynamical symmetry algebra of the
`Schr\"odinger equation' (\ref{inveq}), with auxiliary conditions (\ref{symm-cond}) and the  Schr\"odinger operator}
\BEQ
\mathscr{S} =\frac{2c-\beta}{c-\beta}\left(\partial_{\rho}-\frac{1}{2{\cal M}}\partial^2_y\right).\label{Binveq}
\EEQ
\underline{\em Note}: Formally, comparing the generators of the case {\bf B} treated here and listed in (\ref{gen-casB}),
with those of the case ${\bf A}$ as listed in (\ref{2Ainfinitty}), we
find that they can be turned into each other by going over from the light-cone coordinates (\ref{cone-lumiere})
$\rho,\sigma$ to the variables $\tau=t$ and $v=t+\beta x$ (and setting $\beta=1$)
and also substituting $\Delta \mapsto \delta$ and $\Gamma \mapsto \gamma$.
Alternatively, we might as well simply set $c=0$, which leads to the reduction of variables $\rho\mapsto \tau=t$ and $\sigma\mapsto v=t+\beta x$
but does not  change the Lie algebra. \\
Clearly, the different physical choices of coordinates may be of relevance in specific applications.

\noindent
{\bf Proof:} The commutators were already checked. The dynamical symmetry follows from the commutators
\BEA && [\mathscr{S},{\cal A}^B_n]=\frac{2c-\beta}{c-\beta}\left((n+1)\rho^n\mathscr{S}
        +n(n+1)\rho^{n-1}\left(\frac{(c-\beta)\delta+\gamma}{2c-\beta}-\frac{1}{4}\right)+\frac{n^3-n}{4}\rho^{n-2}{\cal M}y^2\right)\nonumber\\
     && [\mathscr{S},{\cal Y}^B_n]=0, \quad [\mathscr{S},Y_p^y]=\frac{2c-\beta}{c-\beta}{\cal M}(p-\demi)(p+\demi)\rho^{p-3/2}y\label{caseAsym}\nonumber 
\EEA
and $[\mathscr{S},M_0]=0$. Clearly, the generators ${\cal A}^B_{0,\pm 1}, {\cal Y}^B_{n}, Y^y_{\pm \demi}, M_0$
leave the solution space of the eq.~(\ref{inveq}) invariant for $\Delta -\Gamma=\frac{1}{4}$, or equivalently $\gamma= \frac{2c-\beta}{4}+(\beta-c)\delta$.
Central extensions are shown in the appendix to arise only for the Virasoro sub-algebras.
\hfill {\sc q.e.d.}

\section{Representations without time-translation-invariance}

Situations of ``physical ageing'' occur if a statistical system is brought out of equilibrium by a rapid change of one or several physical control parameters.
For example, consider a magnetic system which will order for temperatures $T$ below its critical temperature $T_c>0$.
Consider a quench from high temperatures into the ordered phase with $T<T_c$, which generically has at least two equivalent,
but physically distinct, stationary states. Then
the system will (i) undergo a slow relaxation where (ii) time-translation-invariance is broken and (iii) dynamical scaling holds true.
Relaxational dynamics with these
properties is usually called {\em ageing} \cite{Struik78,Henkel10}.
What are the dynamical symmetries of ageing systems, besides dynamical scaling (generator $X_0$) which is
present as a defining property~? At first sight, and in contrast to equilibrium dynamics, the breaking of time-translation-invariant might suggest that the
generator $X_{-1}=-\partial_t$ of time-translations cannot be part of the symmetry algebra. The dynamical symmetry of ageing systems should then become a sub-algebra
of any dynamical symmetry at equilibrium (which is time-translation-invariant).

However, for the meta-Schr\"odinger algebra, such a  prescription is in general not adequate. This comes about from eq.~(\ref{Bcomhybrid}) which shows
that without the generator $X_{-1}$ of time-translations, the conformal algebra of the generators $Y_n^x$ does not close, especially
\BEQ
[Y^x_0, Y^x_{-1}]=\alpha X_{-1}+\beta Y^x_{-1}.\label{problemcomxminus1}
\EEQ
Obviously, one might either restrict the representations to the trivial solution with $\alpha=0$
or else also exclude the spatial translations $Y^x_{-1}=-\partial_x$ from the
symmetry algebra. Here, we shall not adopt any of those but shall rather ask whether a different construction may be found such that  commutators as
(\ref{problemcomxminus1}), with $\alpha\ne 0$, can be kept.

Indeed, an alternative possibility exists \cite{Minic12,Henkel15}. We rather modify the explicit form of the generator $X_{-1}$, by adding a
time-dependent `potential' term\footnote{We anticipate that spatial translation-invariance, with respect to both $x$ and $y$, are to be kept unchanged.}
\BEQ
X_{-1}= -\partial_t ~~\mapsto~~  X_{-1}^{\rm age}= -\partial_t-V(t).\label{modttg}
\EEQ
Does such a change merely leads to a different representation of the same Lie algebra or can new dynamical algebra be found~?
To answer this, we study the changes implied by (\ref{modttg}) in the  other generators and on the form of the invariant equation.
We shall also impose the physical condition that the dilatation generator must be kept unchanged\footnote{This requirement makes the equivalence with the
time-translation-invariant representations of section~2 a little opaque, see below.} \cite{Henkel06c}
\BEQ
X_0^{\rm age} = X_0 = -t\partial_t-x\partial_x-\frac{y}{2}\partial_y-\delta.\label{scalenonchange}
\EEQ
and should remain in the Cartan sub-algebra. The explicit representations (\ref{modttg},\ref{scalenonchange}) require the
commutator $[X_0,X_{-1}^{\rm age}]\stackrel{!}{=}X_{-1}^{\rm age}$ which in turn fixes the potential, $V(t)=-\xi t^{-1}$, where $\xi$ is a constant.

Next, we can find a generalised invariant equation. In the case of meta-Schr\"odinger symmetry, in the most simple case this will take the form
\BEQ
\mathscr{S}^{\rm age}\Phi(t,x,y)=\bigl(\partial_t-S_1\partial_x-S_2\partial^2_y-\xi t^{-1}\bigr)\Phi(t,x,y)=0.\label{ageinveq}
\EEQ
and the dynamical time-translation and dilatation symmetries are expressed through the commutators
\BEQ
[\mathscr{S}^{\rm age}, X_{-1}^{\rm age}]=0, \quad [\mathscr{S}^{\rm age},X_0]=-\mathscr{S}^{\rm age}\label{simminuszero}
\EEQ
In what follows, the superscript ${}^{\rm age}$ almost always will be dropped. The dynamical symmetry algebra of (\ref{ageinveq}) can be found as in section~2, from the
requirement that the commutation relations (\ref{Bcomhybrid}) and dynamical symmetries (\ref{symmetry}) hold.

\subsection{Ageing representation of the meta-Schr\"odinger algebra: case A}

As before, we take up the two cases (A) $\alpha=0$ and (B) $\alpha\ne 0$. We look at case A first.
Comparing the generators (\ref{Ainfinitty}) of the neat-Schr\"odinger-Virasoro algebra
$\mathfrak{msv}(1,1)$ with the examples (\ref{modttg},\ref{scalenonchange}), it is suggestive that the generators (\ref{Ainfinitty})
should be extended by adding scalar term $F_n(t,x,y)$
and fixing their form from algebraic consistency and from keeping the dynamical symmetry of (\ref{ageinveq}).
It is straightforward to show that an unique solution to these requirements
exists, quite in analogy to the known procedure for the Schr\"odinger-Virasoro algebra.
In terms of the coordinates $t,x,y$, we find that only the generator $X_n$ is changed into
\BEA
X_n &=& - t^{n+1}\partial_t -\frac{1}{\beta}\bigl((t+\beta x)^{n+1} - t^{n+1}\bigr)\partial_x - \frac{n+1}{2}t^n y \partial_y
- \frac{n(n+1)}{4}{\cal M} t^{n-2} y^2
\nonumber \\
& & -(n+1)\frac{\gamma}{\beta}\bigl((t+\beta x)^n - t^n\bigr) - (n+1)\delta t^n - n \xi t^n \label{AgenX1}
\EEA
whereas the other generators maintain their form as stated in (\ref{Ainfinitty}). The only new element is a `second scaling dimension' $\xi$, which together with the
`first scaling dimension' $\delta$ and the `rapidity' $\gamma$ can be used to characterise scaling operators out of stationary states.
Again, it is useful to go over to the time-space coordinates $\tau=t$, $v=t+\beta x$ and $y$. Setting again
$A_n = X_n -\frac{1}{\beta} Y^x_n$, we obtain
\BEA
A_n & = & -{\tau}^{n+1}\partial_{\tau}-\frac{n+1}{2} \tau^n y\partial_y-(n+1)\bigl(\delta-\frac{\gamma}{\beta}\bigr)\tau^n-n\xi \tau^n
          -\frac{n(n+1)}{4}{\cal M}\tau^{n-1}y^2\nonumber\\
Y^x_n & = & -\beta v^{n+1}\partial_v-(n+1)\gamma v^n,\nonumber \\
Y^y_p &=& -\tau^{p+\demi}\partial_y-(p+\demi){\cal M}\tau^{p-\demi}y, \nonumber \\
M_n &=&-{\cal M}\tau^n,\label{2Ainfinitty-xi}
\EEA
Since the commutators (\ref{Acomhybrid2}) all hold true for all $n\in\mathbb{Z}$ and $p\in\mathbb{Z}+\demi$,
the generators (\ref{2Ainfinitty-xi}) span a new representation
of the meta-Schr\"odinger-Virasoro algebra $\mathfrak{msv}(1,1)$.

Turning to the dynamical symmetries of the equation (\ref{ageinveq}), and setting $S_1=\frac{1}{\beta}$,
it is enough to observe that all commutators in (\ref{Asymmetryhybrid}) can be taken over, with the only exception of $[\mathscr{S}^{\rm age},X_1]$, which now reads
\BEQ
[\mathscr{S}^{\rm age},X_1]=-2t\mathscr{S}^{\rm age} +\bigl(2{\cal M}S_2 -1\bigr)y\partial_y -2\bigl(\delta+\xi-\frac{\gamma}{\beta}-\frac{1}{4}\bigr)
\EEQ
Hence the conditions for a symmetry are now $S_2 = \frac{1}{2{\cal M}}$ and $\frac{\gamma}{\beta} = \delta + \xi - \frac{1}{4}$.

\subsection{Ageing representation of the meta-Schr\"odinger algebra: case B$_1$}

Taking up now the case $\alpha\ne 0$, we make use of the known representations as presented in section~2. Wet set again $\alpha=c(c-\beta)$.
In particular, we had seen in the case with time-translation-invariance that all generators can be explicitly written as in (\ref{gen-casB}),
where the light-cone coordinates $\rho,\sigma,y$, see (\ref{cone-lumiere}), are used.
In addition, we see from (\ref{2Ainfinitty-xi}) of case A with $\alpha=0$ that an analogous representation exists for $\alpha\ne0$ if we formally
replace $(\tau,v,y)\mapsto (\rho,\sigma,y)$. With the additional correspondences $\Delta \mapsto \delta$ and $\Gamma\mapsto \gamma$,
the meta-Schr\"odinger generators are completely analogous to the ones written
down in eq.~(\ref{2Ainfinitty-xi}) above.  In terms of the invariant equation, this means that we consider the ageing Schr\"odinger operator
\BEQ
\mathscr{S}^{\rm age}=\frac{2c-\beta}{c-\beta}\left(\partial_{\rho}-\frac{1}{2{\cal M}}\partial^2_y- \frac{\Xi}{\rho}\right)
=\Bigl( \partial_t-\frac{1}{\beta-c}\partial_x-\frac{1}{2{\cal M}}\frac{2c-\beta}{c-\beta}\partial^2_y-\frac{2c-\beta}{c-\beta}\Xi\cdot \frac{1}{t+cx}\Bigr).
\label{ageinveq-xi}
\EEQ
where we now identify $\xi := \frac{2c-\beta}{c-\beta}\Xi$. The conditions on the dynamical symmetry are those derived in section~2, with the only exception
of $\Gamma = \Delta + \Xi -\frac{1}{4}$.

Clearly, an important physical difference with respect to case A is that the form of the potential term in the invariant equation is now of the form $\sim \frac{1}{t+c x}$,
rather than simply $\sim \frac{1}{t}$. Furthermore, the symmetry now acts on $t,x$ in a more subtle way. We shall illustrate this for the  `translation' generators
${\cal X}_{-1}=-\partial_{\rho}+\Xi \rho^{-1}$ and ${\cal Y}^{B}_{-1}=-\partial_{\sigma}$. Using the relations $\partial_{t}=\partial_{\rho}+\partial_{\sigma}$ and
$\partial_x = c\partial_{\rho}+(\beta-c)\partial_{\sigma}$, we have
\BEQ
{\cal X}_{-1} = - \frac{1}{\beta-2c} \left( \bigl(\beta-c\bigr)\partial_t - \partial_x\right) +\frac{\beta-c}{\beta-2c} \frac{\xi}{t+cx} \;\; , \;\;
{\cal Y}^B_{-1} = - \frac{1}{\beta-2c} \bigl( \partial_x - c\partial_t \bigr)
\EEQ
Translations acting on $t,x$ alone are found from
\BEQ
{\cal X}_{-1} + {\cal Y}^B_{-1} = - \partial_t + \frac{\beta-c}{\beta-2c}\frac{\xi}{t+cx} \;\; , \;\;
c {\cal X}_{-1} + (\beta-c){\cal Y}^B_{-1} = -\partial_x +  \frac{c(\beta-c)}{\beta-2c}\frac{\xi}{t+cx}
\EEQ
but they do both contain an extra term which describes the non-trivial transformation of the scaling operator.

\subsection{Ageing representation of the meta-Schr\"odinger algebra: case B$_2$}
Finally, we return to the case where the potential in the invariant equation is of the form $\sim \frac{1}{t}$, but we insist on looking for representations
with $\alpha=c(c-\beta)\ne 0$. First, for the maximal finite-dimensional sub-algebra, we find the explicit generators of $\mathfrak{metasch}(1,1)$
\BEA
X_{-1} & = & -\partial_t+\xi t^{-1}, \quad X_0=-t\partial_t-x\partial_x-\frac{y}{2}\partial_y-\delta\nonumber\\
X_1 & = & -(t^2+\alpha x^2)\partial_t-(2tx+\beta x^2)\partial_x-(t+cx)y\partial_y-(2\delta+\xi)t-2\gamma x-\frac{\cal M}{2}y^2+\alpha\xi t^{-1}x^2\nonumber\\
Y^x_{-1} & = & -\partial_x, \quad Y^x_0=-\alpha x\partial_t-(t+\beta x)\partial_x-\frac{c}{2}y\partial_y-\gamma+\alpha\xi t^{-1}x\nonumber\\
Y^x_1 & = & -\alpha(2tx+\beta x^2)\partial_t-(t^2+2\beta tx+(\alpha+\beta^2)x^2)\partial_x-(ct+(\alpha+\beta c)x)y\partial_y\nonumber\\
      &   & -2\gamma t-2(\alpha\delta+\beta \gamma)x-\frac{c{\cal M}}{2}y^2+\alpha\beta\xi t^{-1}x^2\nonumber\\
Y^y_{\demi} & = & -\partial_y, \quad Y^y_{\demi}=-(t+cx)\partial_y-{\cal M}y, \quad M_0=-{\cal M}.\label{Agemetasch}
\EEA
Here $\alpha, \beta,c,\gamma$ are the parameters of the algebra, with the non-vanishing commutators (\ref{Bcomhybrid}). In applications, their values will
be the same in all observables. In addition, $\delta,\xi$, $\gamma$ and ${\cal M}$
are constants which describe how scaling operators transform under the time-space transformations (\ref{Agemetasch}). Their values can be used to distinguish
different scaling operators. The ageing representation (\ref{Agemetasch}) does contain
the additional scaling dimension $\xi$ which would vanish if time-translation-invariance would be imposed.
The construction of the generators (\ref{Agemetasch}) follows the original prescription \cite{Henkel06c} which requires that the dilatation generator $X_0$ must be kept 
as in the time-translation-invariant case $\xi=0$. 

In order to find the extension to the infinite-dimensional case, we go over to the generators
${\cal A}_n$ and ${\cal Y}_n$. With (\ref{calAdef}) we have $X_{n} = {\cal A}_n + {\cal Y}_n$ and
(\ref{calydef}) gives $Y_n^x = -\bigl( (2\beta-c) {\cal Y}_n - c \bigl( {\cal A}_n + {\cal Y}_n\bigr) \bigr)$.
We give both the explicit expressions in terms of the standard coordinates $(t,x,y)$ and the light-cone coordinates $(\rho,\sigma,y)$ (see (\ref{cone-lumiere}))
\begin{subequations}\label{gen3.13}\begin{align}
{\cal A}_n =
& -\left( t+cx \right) ^{n+1} \frac{ {\left( c-\beta \right) {\partial_t} +  \partial_x} }{2c-\beta}
-\frac{n+1}{2}  \left( t+cx \right)^{n} y\partial_y
-\frac{n  (n+1)}{4}  {\cal M}\left( t+cx \right) ^{n-1} {y}^{2}
\nonumber \\
&  - \left( n+1 \right)  \frac{\gamma+ \left( c-\beta \right) \delta}{2c-\beta} \left( t+cx \right) ^{n}
    +\frac{c-\beta}{2c-\beta} \left( cx-nt \right)  \left( t+cx \right)^{n} \frac{\xi}{t}
\\
= &-\rho^{n+1}\partial_{\rho}-\frac{n+1}{2} \rho^n y\partial_y -\frac{n(n+1)}{4}{\cal M}\rho^{n-1} y^2
\nonumber \\
&  -(n+1)\frac{(c-\beta)\delta+\gamma}{2c-\beta}\rho^n
+ (c-\beta ) \frac{n\bigl( (\beta - c)\rho - c\sigma\bigr) + c(\rho - \sigma)}{(\beta-c)\rho - c\sigma}\rho^n \, \xi
\end{align}\end{subequations}
\begin{subequations}\label{gen3.14}\begin{align}
{\cal Y}_n =
&-\frac {1}{2c-\beta} \bigl(  \left( t+ \left( \beta-c \right) x \right)^{n+1} \left(
 +c\partial_t -\partial_x\right) + \left( n+1 \right)  \left( c\delta-\gamma \right)  \left( t+ \left( \beta-c \right) x \right)^{n}  \bigr)
 \nonumber \\
 &+ {\frac{c\xi}{2c-\beta}\frac{\, \left( -nt+ \left( \beta-c \right) x \right)  \left( t+ \left( \beta-c \right) x \right) ^{n} }{t}}
  \\
=& -\sigma^{n+1}\partial_{\sigma}-(n+1)\frac{c\delta-\gamma}{2c-\beta}\sigma^n
+ c\xi\,{\sigma}^{n}{\frac {     (n+1) \bigl((\rho +\sigma) c\ - \rho\beta\bigr) - (2 c-\beta)\sigma}{ \left(  \left( \rho+\sigma \right) c-\rho\,\beta \right)
 \left( 2c-\beta \right) }}
\end{align}\end{subequations}
which for $\xi=0$ reduce to the explicit forms (\ref{Bcalyafinal},\ref{calBafinal}) and (\ref{calAY-B}), respectively. They obey the commutators (\ref{msv-B}).

Which of these generators lead to dynamical symmetries of the equation (\ref{ageinveq})~? First, consider the generators (\ref{Agemetasch}) of the finite-dimensional
sub-algebra $\mathfrak{metasch}(1,1)$. Working out the commutators
\BEA
&& [\mathscr{S},X_{-1}]=[\mathscr{S},Y^x_{-1}]=[\mathscr{S},Y^y_{-\demi}]=[\mathscr{S},Y^y_{\demi}]=[\mathscr{S},M_0]=0\nonumber\\
&& [\mathscr{S},X_0]=-\mathscr{S}, \quad [\mathscr{S},Y^{x}_0]=-c\mathscr{S}\label{agesymmetry}\\
&& [\mathscr{S},Y^{x}_1]=-2c(t+cx)\mathscr{S}+c\bigl(2{\cal M}S_2 - \frac{2c-\beta}{c-\beta}\bigr)y\partial_y
   +\frac{2c\gamma}{\beta-c}+ c\bigl({\cal M} S_2 - 2\delta - 2\xi\bigr) 
   ,\nonumber\\
&& [\mathscr{S},X_1]=-2(t+cx)\mathscr{S}+\bigl(2{\cal M}S_2 - \frac{2c-\beta}{c-\beta}\bigr)y\partial_y
   +\frac{2\gamma}{\beta-c} + \bigl({\cal M} S_2 - 2\delta - 2\xi\bigr) 
   ,\nonumber
\EEA
dynamical symmetries are obtained if we fix the coefficient $S_{1}=1/(\beta-c)$. Clearly, for the generators $Y_1^x$ and $X_1$ to give a dynamical symmetry
one must fix $S_{2}=\frac{1}{2{\cal M}}\frac{2c-\beta}{c-\beta}$. In addition, reinsertion into  (\ref{agesymmetry}) shows that one must have
$\gamma=\frac{2c-\beta}{4}+(\xi+\delta)(c-\beta)$ as well.
The conditions on $S_1$ and $S_2$ are identical to the requirements in case of time-translation-invariance.

For the infinite-dimensional extension to $\mathfrak{msv}(1,1)$, we use the generators (\ref{gen3.13},\ref{gen3.14}) and also (\ref{Bgalileiinf}).  
With the above choices of $S_1$, $S_2$ and $\gamma$, we then find
\BEA
&& [\mathscr{S},{\cal A}_n] = -(n+1) \bigl( t+ c x\bigr)^n \mathscr{S} - \frac{n^3-n}{4}\frac{\beta+1-2c}{\beta-c}{\cal M} y^2 \bigl( t+ cx\bigr)^n \nonumber \\
&& [\mathscr{S},{\cal Y}_n] = 0 \label{infsymmetry} \\
&& [\mathscr{S},Y_p^y] = - \left( p -\demi\right)\left(p+\demi\right) \frac{\beta-2c}{\beta-c}{\cal M} y \bigl(t+cx\bigr)^{p-3/2} \nonumber \\
&& [\mathscr{S}, M_n] = - n\frac{\beta-2c}{\beta-c} {\cal M} \bigl(t+cx\bigr)^{n-1} \nonumber
\EEA
The conclusions now depend on the choice of the sub-algebra. For the extended Heisenberg algebra
$\left\langle Y_p^y, M_n\right\rangle_{p\in\mathbb{Z}+\frac{1}{2},n\in\mathbb{N}}$, only the generators  of the finite-dimensional sub-algebra
$\mathfrak{hei}(1)=\left\langle Y_{\pm 1/2}^y, M_0\right\rangle$ are dynamical symmetries of the equation
$\mathscr{S}\phi=0$. On the other hand, the meta-conformal sub-algebra {\it alone} $\left\langle {\cal A}_n, {\cal Y}_n\right\rangle_{n\in\mathbb{Z}}$ is
an infinite-dimensional Lie algebra of dynamical symmetries, at least if one chooses $\beta=2c-1$. However, these two sub-algebra cannot be combined into an
infinite-dimensional algebra of dynamical symmetries. Hence the maximal dynamical symmetry algebra is $\mathfrak{vir}\oplus\mathfrak{sch}(1)$, with maximal finite-dimensional sub-algebra $\mathfrak{metasch}(1,1)$.

Of course, an analogous conclusion will hold for any transverse dimension $d_{\perp}\geq 1$.


\subsection{Equivalence of representations}
The `ageing' representations constructed above are equivalent to the corresponding non-ageing representations given in section~2, as we now briefly show. 

\subsubsection{Case A}
The `ageing' Schr\"odinger operator $\mathscr{S}^{\rm age}$ can be obtained from the non-ageing one via the similarity transformation
\BEQ \label{simtransfA}
\mathscr{S}^{\rm age}=\exp(\xi \ln t)\mathscr{S}\exp(-\xi\ln t)=\mathscr{S}-\xi/t 
\EEQ
In principle, the Lie algebra generators should transform similarly. However, because of the requirement $X_0^{\rm age}\stackrel{!}{=}X_0$ taken over from \cite{Henkel06c}, we rather find
\BEA
\wit{X}_n^{\rm age} &:=& \exp(\xi \ln t)X_n\exp(-\xi\ln t) \nonumber \\
&=& - t^{n+1}\partial_t -\frac{1}{\beta}\bigl((t+\beta x)^{n+1} - t^{n+1}\bigr)\partial_x - \frac{n+1}{2}t^n y \partial_y
- \frac{n(n+1)}{4}{\cal M} t^{n-2} y^2
\nonumber \\
& & -(n+1)\frac{\gamma}{\beta}\bigl((t+\beta x)^n - t^n\bigr) - (n+1)\bigl(\delta-\xi\bigr) t^n - n \xi t^n \nonumber \\
&=& \biggl.X_n^{\rm age}\biggr|_{\delta\mapsto \delta-\xi}
\EEA
where $X_n^{\rm age}$ is given in (\ref{AgenX1}). In this way, the `ageing' generator $\wit{X}_n^{\rm age}$, similarity-transformed from  the standard generator (\ref{Ainfinitty}), 
is obtained from (\ref{AgenX1}) by formally substituting $\delta\mapsto \delta-\xi$. Since the other generators do not contain $\delta$, we simply have
\BEQ
\wit{Y}_n^{x,{\rm age}} =Y_n^{x,{\rm age}} = Y_n^x \;\; ,\;\; \wit{Y}_p^{y,{\rm age}}=Y_p^{y,{\rm age}}=Y_p^y \;\; , \;\; \wit{M}_n^{\rm age} = M_n^{\rm age} = M_n
\EEQ

\subsubsection{Case B$_1$}
Because of the mapping of variables $(\tau,v,y)\mapsto (\rho,\sigma,y)$, the similarity transformation is now defined as
\BEQ \label{simtransfB}
\mathscr{S}^{\rm age}=\exp(\xi \ln \rho)\mathscr{S}\exp(-\xi\ln \rho)=\mathscr{S}-\frac{\xi}{\rho} =\mathscr{S}-\frac{\xi}{t+c x} 
\EEQ
The correspondence between the generators follows directly from case A. 

\subsubsection{Case B$_2$}
In this case, the `ageing' Schr\"odinger operator (\ref{ageinveq}) is again produced by the similarity transformation (\ref{simtransfA}). 
Applying this to the similarity transformation of the generators of the meta-Schr\"odinger-Virasoro algebra (\ref{Bcalyafinal},\ref{calBafinal}) and (\ref{calAY-B}), we find 
\BEA  
\wit{\cal Y}_n^{\rm age} & := & \exp(\xi \ln t){\cal Y}_n\exp(-\xi\ln t)\nonumber\\
& = & -\exp(\xi \ln t)\left( t+ (\beta-c)x \right)^{n+1}\frac{c\partial_t -\partial_x}{2c-\beta}\bigl(\exp(-\xi\ln t).\bigr)-(n+1)\frac{c\delta-\gamma}{2c-\beta}\left( t+(\beta-c)x \right)^{n}\nonumber\\
& = & -\left(t+(\beta-c)x\right)^{n+1}\frac{c\partial_t-\partial_x}{2c-\beta}-(n+1)\frac{c\delta-\gamma}{2c-\beta}\left(t+(\beta-c)x\right)^n+\frac{c\xi}{t}
\frac{\left(t+(\beta-c)x\right)^{n+1}}{2c-\beta}\nonumber\\
& = & -\left(t+(\beta-c)x\right)^{n+1}\frac{c\partial_t-\partial_x}{2c-\beta}\nonumber\\
&   & -\left((n+1)(c\delta-\gamma)-c\xi-c\xi(\beta-c)\frac{x}{t}-nc\xi+nc\xi\right)\frac{\left(t+(\beta-c)x\right)^{n}}{2c-\beta}\nonumber\\
& = & -\left(t+(\beta-c)x\right)^{n+1}\frac{c\partial_t-\partial_x}{2c-\beta}-(n+1)(c(\delta-\xi)-\gamma)\frac{\left(t+(\beta-c)x\right)^{n}}{2c-\beta}\nonumber\\
&   & +(-nt+(\beta-c)x)\frac{c\xi}{t}\frac{\left(t+(\beta-c)x\right)^{n}}{2c-\beta}\label{ynsimtransf}\\
&=& \biggl. {\cal Y}_n^{\rm age}\biggr|_{\delta\mapsto \delta-\xi}
\nonumber
\EEA
where ${\cal Y}_n^{\rm age}$ is given by (\ref{gen3.14}). Similarly 
\BEA
\wit{\cal A}_n^{\rm age} & := & \exp(\xi \ln t){\cal A}_n\exp(-\xi\ln t)\nonumber\\
& = & -\left( t+cx \right) ^{n+1} \frac{ {\left( c-\beta \right) {\partial_t} +  \partial_x} }{2c-\beta}
-\frac{n+1}{2}  \left( t+cx \right)^{n} y\partial_y
-\frac{n  (n+1)}{4}  {\cal M}\left( t+cx \right) ^{n-1} {y}^{2}
\nonumber \\
&   & - \left( n+1 \right)  \frac{\gamma+(c-\beta)(\delta-\xi)}{2c-\beta}\left( t+cx \right) ^{n} +
\frac{c-\beta}{2c-\beta} \left( cx-nt \right)  \left( t+cx \right)^{n} \frac{\xi}{t}\nonumber\\
&=& \biggl. {\cal A}_n^{\rm age}\biggr|_{\delta\mapsto \delta-\xi}
\label{ansimtransf}
\EEA
where ${\cal A}_n^{\rm age}$ is given by (\ref{gen3.13}). As before the  effect of the similarity transformation reduces to the formal substitution $\delta\mapsto\delta - \xi$ in the 
`first' scaling dimension. Finally, the generators $Y_p^y$ and $M_n$ do not contain $\delta$ and are therefore invariant, viz. 
\BEQ
\wit{Y}_p^{y,{\rm age}}=e^{\xi\ln t}\, Y_p^y\, e^{-\xi\ln t} = Y_p^y \;\; , \;\; \wit{M}_n^{\rm age}=e^{\xi\ln t}\, M_n\, e^{-\xi\ln t}=M_n
\EEQ

\section{Covariant two-point functions}

As an application, we consider the form of the co-variant two point-function, defined as
\BEQ
F(t_1,t_2,x_1,x_2,y_1,y_1)=\bigl\langle \Phi_1(t_1,x_1,y_1)\Phi_2(t_2,x_2,y_2)\bigr\rangle\label{deftwo}
\EEQ
where $\Phi_1,\Phi_2$ are quasi-primary scaling operators of $\mathfrak{metasch}(1,1)$. From the representations constructed before,
the scaling operators are characterised by the several parameters introduced.
Physically, we distinguish the {\em stationary case}, which is time-translation-invariant and the {\em ageing case}, which is not.

\subsection{Stationary case}
We begin with time-translation-invariant case.
Since all representations are equivalent, we consider scaling operators $\Phi_{i}$ ($i=1,2$) transforming covariantly under the
representation (\ref{Ametasch}) of \underline{case A}.
Scaling operators are characterised by the parameters $(\delta_i, \frac{\gamma_i}{\beta_i}, {\cal M}_i, \beta_i)$.
Lifting the single-body representation (\ref{Ametasch}) to a two-body representation, we obtain the following Ward identities:
\begin{enumerate}
\begin{subequations} \label{Acov}
\item The action of the time- and space-translations $X_{-1},Y^x_{-1},Y^y_{-\demi}$ on (\ref{deftwo}) obviously leads to
\BEQ
F= F(t,x,y) \;\; ; \;\; \mbox{\rm\small with $t=t_1-t_2$, $x=x_1-x_2$ and $y=y_1-y_2$}
\EEQ
\item The action of the dilatation-generator $X_0$ produces:
\BEQ \label{AcovX0}
\bigl(-t\partial_t-x\partial_x-\demi y\partial_y-\delta_1-\delta_2\bigr)F(t,x,y)=0
\EEQ
\item The action of the generalised `parallel' Galilei transformation $Y_0^x$ gives
\BD
\bigl(  - t\partial_x - (\beta_1 x_1-\beta_2 x_2)\partial_x - \gamma_1 - \gamma_2 \bigr)F(t,x,y) = 0
\ED
which we can re-write as $\bigl(-t\partial_x-\bigl[ \beta_1 (x_1-x_2) + (\beta_1-\beta_2)x_2\bigr]\partial_x -\gamma_1 - \gamma_2\bigr)F=0$.
Because of spatial `parallel' translation-invariance, no terms containing explicitly the coordinate $x_2$ are admissible. Therefore
\BEQ \label{AcovY0}
\bigl( -(t + \beta_1 x)\partial_x - \gamma_1 - \gamma_2 \bigr)F(t,x,y) = 0 \;\; ; \;\; (\beta_1-\beta_2) x_2 \partial_x F(t,x,y) = 0
\EEQ
which in particular implies the constraint $\beta_1=\beta_2 =: \beta$. The velocity $\beta^{-1}$ arises in the Schr\"odinger
operator (\ref{Ainveq}) and since it is the same for any pair of scaling operators, it must be an universal constant.
\item The action of the generator $Y_1^x$ gives
\BD
\left\{ -\bigl[ (t_1+\beta x_1)^2 -(t_2+\beta x_2)^2\bigr]\partial_x -2\gamma_1 (t_1+\beta x_1) -2\gamma_2 (t_2 +\beta x_2)\right\}F(t,x,y) = 0
\ED
Re-using the identity (\ref{AcovY0}), this can be reduced to
\BEQ \label{AcovY1}
\bigl( -(t + \beta x)^2\partial_x - 2\gamma_1 (t+\beta x) \bigr)F(t,x,y) = 0
\EEQ
and comparing this again with (\ref{AcovY0}) leads to the constraint $\gamma_1 = \gamma_2$.
\item The action of the `transverse' Galilei transformation $Y^y_{\demi}$ produces
\BD
\bigl( -t\partial_y - {\cal M}_1 y_1 - {\cal M}_2 y_2 \bigr) F(t,x,y) = 0
\ED
which is re-written as $\bigl( -t\partial_y - {\cal M}_1 (y_1 -y_2)- ({\cal M}_1 + {\cal M}_2) y_2 \bigr) F(t,x,y) = 0$. Again, because of
spatial `transverse' translation-invariance, an explicit occurrence of the coordinate $y_2$ is inadmissible. Hence
\BEQ \label{AcovY1/2}
\bigl( -t\partial_y - {\cal M}_1 y \bigr) F(t,x,y) = 0 \;\; ; \;\; \bigl({\cal M}_1 + {\cal M}_2\bigr) y_2 F(t,x,y) = 0
\EEQ
which in particular leads to the constraint ${\cal M}_2 = - {\cal M}_1$.
\item Finally, we must consider the action of the generator $X_1$. The rather lengthy expressions can be simplified as before, and re-using the
conditions (\ref{AcovX0},\ref{AcovY0},\ref{AcovY1/2}). This leads to the further constraint $\delta_1=\delta_2$.
\end{subequations}
\end{enumerate}
It follows that the Ward identities (\ref{Acov}) fix the form of the two-point function,
through the three differential equations (\ref{AcovX0},\ref{AcovY0},\ref{AcovY1/2}).
Setting $F(t,x,y) = G^{(A)}(t,t+\beta x,y)$, and letting $v=t+\beta x$,
standard techniques directly lead to, with $\beta_1=\beta_2=\beta$
\BEQ
G^{(A)}(t,v,y) =  G_0\, \delta_{{\cal M}_1+{\cal M}_2,0}\:\delta_{\delta_1,\delta_2}\delta_{\beta_1,\beta_2}\delta_{\gamma_1,\gamma_2}\;
t^{-2\delta_1+{2\gamma_1}/{\beta}}\, v^{-2{\gamma_1}/{\beta}} \exp\left(-\frac{{\cal M}_1}{2}\frac{y^2}{t}\right).\label{Adeuxpointsfinal}
\EEQ
and the normalisation constant $G_0$. This form combines aspects of Schr\"odinger-invariance in the transverse coordinate $y$ and of
meta-conformal invariance in the parallel coordinate $x$.

Turning to \underline{case B}, we have already seen that this corresponds to different representations of the same algebra. This is obvious by comparing the
generators (\ref{2Ainfinitty}) which are formulated in the coordinates $(\tau,v,y)=(t,t+\beta x,y)$, with the generators (\ref{gen-casB})
which are formulated in the coordinates $\rho=t+cx$ and $\sigma=t+(\beta-c)x$ and also use the definitions
(\ref{defGammaDelta}).\footnote{Now, both $c$ and $\beta-c$ are universal constants.}
Then eqs.~(\ref{Acov}) are readily adapted. The end result is, where $F(t,x,y) = G^{(B)}(\rho,\sigma,y)$
\BEQ
G^{(B)}(\rho,\sigma,y) =  G_0\, \delta_{{\cal M}_1+{\cal M}_2,0}\:\delta_{\beta_1,\beta_2}\delta_{c_1,c_2}\delta_{\Delta_1,\Delta_2}\delta_{\Gamma_1,\Gamma_2}\;
\rho^{-2\Delta_1+{2\Gamma_1}}\, \sigma^{-2{\Gamma_1}} \exp\left(-\frac{{\cal M}_1}{2}\frac{y^2}{\rho}\right) \label{Bdeuxpointsfinal}
\EEQ
and where we also have $\rho = \rho_1-\rho_2$, $\sigma=\sigma_1-\sigma_2$ and $y=y_1-y_2$.

\subsection{Ageing case}

For non-equilibrium dynamics, as it occurs in ageing systems, time-translation-invariance does not hold and the scaling operators
$\Phi_{1,2}$ are quasi-primary with respect to the representations derived in section~3.

Again, we consider first \underline{case A} and have the modified representations (\ref{AgenX1},\ref{2Ainfinitty-xi}).
The time variables are now $t=t_1-t_2$ and $u =t_1/t_2$. Most of the treatment from the stationary case can be taken over, but there are some important modifications.
\begin{subequations} \label{Acov-xi}
\begin{enumerate}
\item The actions of the spatial translations $Y_{-1}^x$ and $Y_{-\demi}^y$ are unchanged, such that
\BEQ
F= F(t,u,x,y) \;\; ; \;\; \mbox{\rm\small with $t=t_1-t_2$, $u=t_1/t_2$, $x=x_1-x_2$ and $y=y_1-y_2$}
\EEQ
but the action of $X_{-1}$ now leads to the condition $\bigl(-\partial_{t_1} - \partial_{t_2} + \xi_1/t_1 + \xi_2/t_2\bigr)F(t,u,x,y)=0$, or equivalently
\BEQ \label{AcovX-1-xi}
\bigl( (u-1)\partial_u + \frac{1}{u} \xi_1 + \xi_2 \bigr)F(t,u,x,y) = 0
\EEQ
\item The action of the dilatation generator $X_0$ finally reproduces (\ref{AcovX0})
\BEQ \label{AcovX0-xi}
\bigl(-t\partial_t-x\partial_x-\demi y\partial_y-\delta_1-\delta_2\bigr)F(t,u,x,y)=0
\EEQ
\item Since the actions of the generators $Y_{0,1}^x$ and $Y_{\demi}^y$ do not imply changes in the time-coordinates $t_{1,2}$, the arguments leading
to eqs.~(\ref{AcovY0},\ref{AcovY1/2}) and the constraints $\gamma_1=\gamma_2$ and ${\cal M}_1+{\cal M}_2=0$ (along with $\beta_1=\beta_2=\beta$)
apply to the present case as well, with $F=F(t,u,x,y)$.
\item The action of $X_1$ must be re-analysed. Using (\ref{AcovX0-xi}) and (\ref{AcovY0},\ref{AcovY1/2}), we readily find
\begin{align}
X_1 F &= \Big\{ -t^2\partial_t -2tx \partial_x -\beta x^2\partial_x -ty\partial_y -\frac{{\cal M}_1}{2}y^2 -2\gamma_1 x -2\delta_1 -\xi_1 t  \nonumber \\
&~~~  +t_2\bigl[ u(1-u)\partial_u -\xi_1 - \xi_2 \bigr] \Big\} F(t,u,x,y) \nonumber \\
&= \Bigl\{ -(\delta_1 - \delta_2) t +t_2\bigl[ u(1-u)\partial_u -\xi_1 - \xi_2 \bigr] \Big\} F(t,u,x,y) \nonumber \\
&= \Bigl\{ -(\delta_1 - \delta_2) t - \xi_1 t + t_2 \bigl[ \xi_1 + u \xi_2 - \xi_1 -\xi_2 \bigr] \Big\} F(t,u,x,y) \nonumber \\
&= -t\bigl( \delta_1 + \xi_1 - \delta_2 - \xi_2\bigr) F(t,u,x,y) = 0
\end{align}
where in the second line, the contributions proportional to (\ref{AcovX0-xi},\ref{AcovY0},\ref{AcovY1/2}) were subtracted off and the constraint
$\gamma_1=\gamma_2$ was used. In the third line, eq.~(\ref{AcovX-1-xi}) was used.  We read off the constraint $\delta_1 +\xi_1 = \delta_2 + \xi_2$.
\end{enumerate}
\end{subequations}
In consequence, the Ward identities (\ref{Acov}) and (\ref{Acov-xi}), respectively, give the form of the quasi-primary two-point function.
Writing $F(t,u,x,y) =G^{(A)}(t,u,v,y)$ with $v=t+\beta x$, we promptly find
\BEA
G^{(A)}(t,u,v,y) &=& G_0 \delta_{{\cal M}_1+{\cal M}_2,0}\: \delta_{\beta_1,\beta_2}\delta_{\gamma_1,\gamma_2} \delta_{\delta_1+\xi_1,\delta_2+\xi_2}\;
t^{-\delta_1-\delta_2+2\gamma_1/\beta} \nonumber \\
& & \times~ u^{\xi_1} (u-1)^{-\xi_1-\xi_2} v^{-2\gamma_1/\beta}\: \exp\left( -\frac{{\cal M}_1}{2}\frac{y^2}{t} \right)
\label{resfinA}
\EEA

We now turn to the ageing representations of case B. We have seen in section~3 that two sub-cases should be distinguished: \\
\underline{\bf (i)} in the \underline{case B$_1$},
all stationary generators formally retain their form with respect to case A  if one merely replaces $(\tau,v,y)\mapsto (\rho,\sigma,y)$.
In the non-stationary case at hand, we just set $u=\rho_1/\rho_2$. With the replacements considered above we simply adapt (\ref{resfinA}). Writing
$F(t_1-t_2,\frac{t_1+c x_1}{t_2+c x_2},x,y)=G^{(B_1)}(\rho,u,v,y)$, we have
\BEA
G^{(B_1)}(\rho,u,v,y) &=& G_0 \delta_{{\cal M}_1+{\cal M}_2,0}\: \delta_{\beta_1,\beta_2}\delta_{\Gamma_1,\Gamma_2} \delta_{\Delta_1+\Xi_1,\Delta_2+\Xi_2}\;
\nonumber \\
& & \times~
\rho^{-\Delta_1-\Delta_2+2\Gamma_1} u^{\Xi_1} (u-1)^{-\Xi_1-\Xi_2} \sigma^{-2\Gamma_1}\: \exp\left( -\frac{{\cal M}_1}{2}\frac{y^2}{\rho} \right)
\label{resfinB1}
\EEA
\underline{\bf (ii)} in the \underline{case B$_2$}, such an adaption is not possible and we must consider the representation (\ref{Agemetasch}).
The Ward identities are then derived as follows.
\begin{subequations} \label{Bcov}
\begin{enumerate}
\item The actions of the spatial translations $Y_{-1}^x$ and $Y_{-\demi}^y$ are unchanged, such that
\BEQ
F= F(t,u,x,y) \;\; ; \;\; \mbox{\rm\small with $t=t_1-t_2$, $u=t_1/t_2$, $x=x_1-x_2$ and $y=y_1-y_2$}
\EEQ
and the action of $X_{-1}$ leads to the condition
\BEQ \label{BcovX-1-xi}
\bigl( u(u-1)\partial_u  + \xi_1 + u \xi_2 \bigr)F(t,u,x,y) = 0
\EEQ
\item The action of the dilatation generator $X_0$ reproduces (\ref{AcovX0-xi})
\BEQ \label{BcovX0-xi}
\bigl(t\partial_t+x\partial_x+\demi y\partial_y+\delta_1+\delta_2\bigr)F(t,u,x,y)=0
\EEQ
\item Covariance under the Galilei generator $Y_{\demi}^y$, again together with the requirements of spatial translation-invariance in both $x$ and $y$, gives
\BEQ \label{BcovY1/2}
\bigl((t+c_1x)\partial_y+{\cal M}_1y\bigr)F(t,u,x,y)=0
\EEQ
together with the constraints $c_1=c_2$ and ${\cal M}_1+{\cal M}_2=0$. The last condition also follows from the covariance under $M_0$.
\item The covariance under the generator $Y^x_0$ is more difficult to analyse. First, the reduction to the variables at hand and the known
constraints give
\begin{align}
&   \left\{ - \alpha_1 x \partial_t - \bigl(t+\beta_1 x\bigr)\partial_x - \frac{c_1}{2}y \partial_y - \gamma_1 - \gamma_2
    - \frac{\alpha_1 x}{t_2} \left( \partial_u - \frac{\xi_1}{u}\right)  
\right.\nonumber \\
&  \left. - \frac{x_2}{t_2} \left( \alpha_1 - \alpha_2 u \partial_u
-\frac{\alpha_1\xi_1}{u} - \alpha_1\xi_2\right)  -\bigl(\alpha_1 - \alpha_2\bigr) x_2 \partial_t - x_2 \bigl(\beta_1-\beta_2\bigr)\partial_x \right\} F=0
\nonumber \tag{*}
\end{align}
Next, using again the requirement of spatial translation-invariance, the last two terms in the second line of (*) give the constraints
$\alpha_1=\alpha_2$ and $\beta_1=\beta_2$.
Since we had already seen above that $c_1=c_2$, this is consistent with the chosen parametrisation $\alpha=c(c-\beta)$.
Third, re-using (\ref{BcovX-1-xi}) implies that the first term in the second line of (*) also vanishes. Forth, we can express
$t_2 = t/(u-1)$. This implies that (*) can be re-written as
\begin{align}
&   \left\{ - \alpha_1 x \partial_t - \bigl(t+\beta_1 x\bigr)\partial_x - \frac{c_1}{2}y \partial_y - \gamma_1 - \gamma_2
    - \frac{\alpha_1 x}{t} \left( (u-1)\partial_u - \xi_1\frac{u-1}{u}\right)   \right\} F=0 \nonumber \tag{**}
\end{align}
which upon re-using again (\ref{BcovX-1-xi}), in the last term, reduces to
\BEQ \label{BcovY0}
\left\{ - \alpha_1 x \partial_t - \bigl(t+\beta_1 x\bigr)\partial_x - \frac{c_1}{2}y \partial_y - \gamma_1 - \gamma_2
    - \alpha_1\frac{x}{t} \bigl( \xi_1 + \xi_2\bigr)   \right\} F=0
\EEQ
\item Finally, straightforward but tedious calculations also show that the action of the generators $Y_1^x$ and $X_1$ imply the
constraints $\gamma_1=\gamma_2$ and $\delta_1+\xi_1=\delta_2+\xi_2$.
\end{enumerate}
\end{subequations}
Summarising,  we have the system of equations (\ref{Bcov}) 
for the function $F(t,u,x,y)$
\begin{subequations} \label{eqs}
\begin{align}
\bigl(u(u-1)\partial_u+\xi_1+u\xi_2)F(t,u,x,y)\bigr) &=0,\label{1eq1}\\
\bigl(t\partial_t+x\partial_x+\frac{y}{2}\partial_y+\delta_1+\delta_2\bigr)F(t,u,x,y)&=0\label{2eq2}\\
\bigl(\alpha_1x\partial_t+(t+\beta_1x)\partial_x+\demi c_1y\partial_y+\gamma_1+\gamma_2+\alpha_1\frac{x}{t}\bigl(\xi_1+\xi_2\bigr)\bigr)F(t,u,x,y)&=0\label{3eq3}\\
\bigl((t+c_1x)\partial_y+{\cal M}_1y\bigr)F(t,u,x,y)&=0.\label{4eq4}
\end{align}
\end{subequations}
where the condition $\alpha=c(c-\beta)$ must be kept.
The solution of (\ref{eqs}) is readily found, where for simplicity of notation we write $\beta=\beta_1=\beta_2$ and $c=c_1=c_2$
\BEA
\lefteqn{F(t,u,x,y) =  F_0 \delta_{{\cal M}_1+{\cal M}_2,0}\delta_{\delta_1+\xi_1,\delta_2+\xi_2}\delta_{\gamma_1,\gamma_2}\delta_{\beta_1,\beta_2}\delta_{c_1,c_2}\,
t^{-\xi_1-\xi_2}u^{\xi_1}(u-1)^{-\xi_1-\xi_2}} \label{resfinal} \\
& \times & (t+(\beta-c)x)^{\frac{c(\delta_1+\delta_2-\xi_1-\xi_2)-\gamma_1-\gamma_2}{\beta-2c}}
           (t+cx)^{\frac{-(\beta-c)(\delta_1+\delta_2-\xi_1-\xi_2)+\gamma_1+\gamma_2}{\beta-2c}}
\exp\left(-\frac{{\cal M}_1}{2}\frac{y^2}{t+cx}\right). \nonumber
\EEA
In terms of the light-cone coordinates $\rho$ et $\sigma$ and letting $F(t,u,x,y)=G^{(B_2)}(t,u,\rho,\sigma,y)$
with $u=\frac{t_1}{t_2}=\frac{(c-\beta)\rho_1+c\sigma_1}{(c-\beta)\rho_2+c\sigma_2}$ we have
\BEA
\lefteqn{G^{(B_2)}(t,u,\rho,\sigma,y) =  G_0
\delta_{{\cal M}_1+{\cal M}_2,0}\delta_{\Delta_1-\Delta_2,-\frac{c}{c-\beta}(\Xi_1-\Xi_2)}\delta_{\frac{2c-\beta}{c}(\Gamma_1-\Gamma_2),-\frac{c}{c-\beta}(\Xi_1-\Xi_2)}
\delta_{\beta_1,\beta_2}\delta_{c_1,c_2}\, }  \nonumber\\
& \times &
\left( \frac{(c-\beta)\rho +c\sigma}{2c-\beta}\right)^{-\frac{2c-\beta}{c-\beta}(\Xi_1+\Xi_2)}
\sigma^{-\Gamma_1-\Gamma_2+\frac{c}{c-\beta}(\Xi_1+\Xi_2)}\,
\rho^{-\Delta_1-\Delta_2+\Xi_1+\Xi_2+\Gamma_1+\Gamma_2}
\nonumber \\
& \times &
u^{\frac{2c-\beta}{c-\beta}\Xi_1}(u-1)^{-\frac{2c-\beta}{c-\beta}(\Xi_1+\Xi_2)}\, \exp\left(-\frac{{\cal M}_1}{2}\frac{y^2}{\rho}\right) \label{resfinalGB2}
\EEA
where we re-used the abbreviations (\ref{defGammaDelta}) and recall the definition $\xi_i = \frac{2c-\beta}{c-\beta}\Xi_i$.
In contrast to the other cases, there is no simple reduction to simple powers of the light-cone coordinates $\rho=t+cx$ and $\sigma=t+(\beta-c)x$,
but a further explicit dependence on the time-difference $t=\frac{(c-\beta)\rho +c\sigma}{2c-\beta}$ is maintained.

Although the functional form of the three solutions (\ref{resfinA},\ref{resfinB1},\ref{resfinalGB2}) for the co-variant two-point functions
are quite similar in the overall shape, the precise dependence on the various scaling dimension as well as the constraints on them are different.

The generalisation of the results (\ref{Adeuxpointsfinal},\ref{Bdeuxpointsfinal},\ref{resfinA},\ref{resfinB1},\ref{resfinalGB2})
to more than one transverse dimension $d_{\perp}=d-1\geq 1$ is achieved by the substitution
$y\mapsto \vec{y}\in\mathbb{R}^{d-1}$.

\section{Conclusions}

The interpolation between the dynamical symmetries of the diffusion equation and of the equation of simple ballistic transport leads to a new class of
symmetries if the associated dynamical exponent is allowed to take different values in dependence of the spatial direction. These have been constructed as time-space
symmetries where several different ways of the action of the time-transformations were considered. Restricting for simplicity to the case of a single
`parallel' and a single `transverse' direction, the infinite-dimensional {\it `meta-Schr\"odinger-Virasoro'} Lie algebra has been obtained,
which is isomorphic to
\BEQ \label{LieA}
\mathfrak{msv}(1,1)= \bigl(\mathfrak{vir}\oplus\mathfrak{vir}\bigr)\ltimes\mathfrak{gal}(1)
                   =       \mathfrak{vir}\oplus\mathfrak{sv}(1)
\EEQ
Its maximal finite-dimensional sub-algebra is the {\it `meta-Schr\"odinger algebra'} $\mathfrak{metasch}(1,1)$.
We generalised these to representations of $\mathfrak{msv}(1,1)$ capable of describing non-stationary dynamical symmetries where time-translation-invariance
does no longer hold true. Then the usual generator $-\partial_t$ of time-translations is replaced by $X_{-1} = - \partial_t + \xi/t$,
where the parameter $\xi$ describes new properties of non-stationary scaling operators. We construct the representations of
$\mathfrak{msv}(1,1)$ with $\xi\ne 0$. The representations of $\mathfrak{metasch}(1,1)\subset\mathfrak{vir}\oplus\mathfrak{sch}(1)\subset\mathfrak{msv}(1,1)$
act as dynamical symmetries of simple Schr\"odinger equations $\mathscr{S}\Phi(t,x,y)=0$, with
\BEQ \label{5.2}
\mathscr{S} = \left\{
\begin{array}{ll}
\partial_t - \frac{1}{\beta}\partial_x - \frac{1}{2{\cal M}} \partial_y^2 - \frac{\xi}{t} & \mbox{\rm case A} \\[0.1truecm]
\partial_t + \frac{1}{c-\beta}\partial_x - \frac{1}{2{\cal M}}\frac{2c-\beta}{c-\beta}\partial_y^2 - \frac{\xi}{t+cx} & \mbox{\rm case B$_1$}
\\[0.1truecm]
\partial_t + \frac{1}{c-\beta}\partial_x - \frac{1}{2{\cal M}}\frac{2c-\beta}{c-\beta} \partial_y^2 -\frac{\xi}{t} & \mbox{\rm case B$_2$}
\end{array} \right.
\EEQ
provided the rapidity $\gamma$ is chosen accordingly. 
These are not time-translation-invariant but the time-translation-invariant case is recovered by setting $\xi=0$.

The representations with $\xi\ne0$ are equivalent to those with $\xi=0$; physical considerations will dictate the choice of the appropriate representation, for example
through the required form of the Schr\"odinger operator (\ref{5.2}). 

The representations of $\mathfrak{metasch}(1,1)$ permit to derive the scaling form of the quasi-primary (co-variant) two-point functions.
These are conveniently written in terms of natural light-cone coordinates $\tau=t$ and $v=t+ \beta x$ or $\rho=t+c x$ and $\sigma=t+(\beta-c) x$, respectively.
Comparing the stationary forms (\ref{Adeuxpointsfinal},\ref{Bdeuxpointsfinal}) with the non-stationary ones (\ref{resfinA},\ref{resfinB1},\ref{resfinalGB2}),
the great overall similarity of the resulting scaling forms is evident, up to a suitable identification of exponents.
In the non-stationary case, the main change is the modification of the form of the scaling function
through a uniquely determined function of the new scaling variable $u$ which depends on the ratio of the two times.
While this is the only change in the more simple cases $A$ and $B_1$,
in the more elaborate non-stationary case $B_2$ a further non-trivial dependence on the time-difference is maintained as well.

A sequel paper will explore physical realisations of these symmetries in the non-equilibrium ageing of spin systems with both diffusive and ballistic transport.

\noindent
{\bf  Acknowledgements:} SS and MH  thank the PHC Rila(KP-06-RILA/7) for financial support. 
SS is also supported by Bulgarian National Science Fund Grant KP-06-N28/6. 
MH is also supported by the french ANR-PRME UNIOPEN. 

\appsektion{On central extensions}

Our discussion on possible central extensions of the meta-Schr\"odinger-Virasoro algebra is based on the following simple observation.

\noindent
{\bf Proposition A.1}: {\it Consider the following infinite-dimensional Lie algebra, with generators $A_n$ with $n\in\mathbb{Z}$ and
$B_k$ with $k\in\mathbb{Z}$ (or $k\in \mathbb{Z}+\demi$). The commutators are}
\BEA
{} \bigl[ A_n , A_m \bigr] &=& \alpha_{n,m} A_{n+m} + c_A(n,m) \nonumber \\
{} \bigl[ A_n , B_k \bigr] &=& d(n,k) \label{A1} \\
{} \bigl[ B_k , B_{\ell} \bigr] &=& \beta_{k,\ell} B_{k+\ell} \hspace{0.35truecm}+ c_B(k,\ell) \nonumber
\EEA
{\it such that $c_A(n,m)$, $c_B(k,\ell)$ and $d(n,k)$ are all central terms.
If the structure constants are such that $\alpha_{n,0}\ne 0$ and $\alpha_{n,-n}\ne 0$ for all $n\ne 0$, then $d(n,k)=0$ for all $n,k\in\mathbb{Z}$
(or $k\in \mathbb{Z}+\demi$).}

Implicitly, the structure constants $\alpha_{n,m}$ and central extension $c_A(n,m)$ are admitted such that the Jacobi identities for the sub-algebra
$\left\langle A_n\right\rangle_{n\in\mathbb{Z}}$ are obeyed, and analogously for $\left\langle B_k\right\rangle_{k}$.

\noindent
{\bf Proof:} Let $n,m\in\mathbb{Z}$, and either $k\in \mathbb{Z}$ or $k\in\mathbb{Z}+\demi$. Then consider the Jacobi identity
$\bigl[ A_n, \bigl[ A_m, B_k \bigr] \bigr] + \bigl[ B_k, \bigl[ A_n, A_m \bigr] \bigr] + \bigl[ A_m, \bigl[ B_k, A_n \bigr] \bigr] = 0$ which gives from (\ref{A1})
\BD
\bigl[ A_n , d(m,k) \bigr] + \bigl[ B_k , \alpha_{n,m} A_{n+m} +c_A(n,m)\bigr] + \bigl[ A_m, - d(n,k) \bigr] = 0
\ED
Since $d$ is central, the first and third brackets vanish. Since $c_A$ is central, it follows
\BD
\alpha_{n,m}\, d(n+m,k) = 0
\ED
Set first $m=0$. Since $\alpha_{n,0}\ne 0$ was admitted, the assertion $d(n,k)=0$ follows for $n\ne 0$.
If alternatively one sets $m=-n\ne 0$ and since one has $\alpha_{n,-n}\ne 0$, it
follows $d(0,k)=0$. This proves the assertion. \hfill {\sc q.e.d.}

The centre-less form of the meta-Schr\"odinger-Virasoro algebra
$\mathfrak{msv}(1,1)\cong\left\langle {\cal A}_n, {\cal Y}_n, Y_{k}^y, M_n\right\rangle_{n\in\mathbb{Z}, k\in\mathbb{Z}+\demi}$ obeys
the commutators (\ref{msv-B}). It can be written in the form of a direct sum
\BEQ \label{A2}
\mathfrak{msv}(1,1) \cong \mathfrak{vir} \oplus \mathfrak{sv}(1) \cong \left\langle {\cal Y}_n \right\rangle_{n\in\mathbb{Z}} \oplus
\left\langle {\cal A}_n, Y_{k}^y, M_n\right\rangle_{n\in\mathbb{Z}, k\in\mathbb{Z}+\demi}
\EEQ
First, the central extensions of the Schr\"odinger-Virasoro algebra $\mathfrak{sv}(1)$ are well-known \cite{Henkel94,Unterberger06,Unterberger12}.
Only the Virasoro sub-algebra $\mathfrak{vir}\cong \left\langle {\cal A}\right\rangle_{n\in\mathbb{Z}}$ admits a central extension, of the familiar form \cite{Kac87}.
All other central terms in the commutators of $\mathfrak{sv}(1)$ either vanish or can be absorbed into the generators. Second, the
sub-algebra $\left\langle {\cal A}_n, {\cal Y}_n\right\rangle_{n\in\mathbb{Z}} \cong \mathfrak{vir}\oplus\mathfrak{vir}$ is but the familiar double
Virasoro algebra. Third, from the explicit commutator (\ref{msv-B}) read off $\alpha_{n,m}=n-m$, such that we have $\alpha_{n,0}=n$ and $\alpha_{n,-n}=2n$
so that we can apply Proposition A.1. 
Finally, because of the direct-sum structure of the commutators of (\ref{A2}), there cannot exist any central term in the
commutators of ${\cal Y}_n$ with any other family of generators in $\mathfrak{msv}(1,1)$. We therefore have

\noindent
{\bf Proposition A.2.} {\it The non-vanishing commutators, including the possible central extensions, of the meta-Schr\"odinger-Virasoro algebra
$\mathfrak{msv}(1,1)$ have the form}
\BEA
{} && \bigl[ {\cal A}_n , {\cal A}_m \bigr] = (n-m) {\cal A}_{n+m} + \frac{c_{\cal A}}{12} \bigl( n^3 - n\bigr) \delta_{n+m,0}
\nonumber \\
{} && \bigl[ {\cal Y}_n , {\cal Y}_m \bigr] = (n-m) {\cal Y}_{n+m} \hspace{0.14truecm}+ \frac{c_{\cal Y}}{12} \bigl( n^3 - n\bigr) \delta_{n+m,0}
\label{A3} \\
{} && [{\cal A}_n,Y^y_k]=\left(\frac{n}{2}-k\right)Y^y_{n+k},\quad~  [Y^y_k, Y^y_{\ell}]=(k-\ell)M_{k+\ell},  \quad [{\cal A}_n,M_m]=-m M_{n+m}~~
\nonumber
\EEA
{\it with $n,m\in \mathbb{Z}$ and $k,\ell\in\mathbb{Z}+\demi$ and the central charges $c_{\cal A}$ and $c_{\cal Y}$ are constants.}

In this purely algebraic discussion, $c_{\cal A}$ and $c_{\cal Y}$ remain independent,
but further considerations such as unitarity and/or hermiticity might provide further constraints. The extension of (\ref{A3}) to the higher-dimensional algebra
$\mathfrak{msv}(1,d_{\perp})$ with $d_{\perp}\geq 1$ is obvious.


\end{document}